\documentclass[prb,aps,twocolumn,showpacs,floatfix]{revtex4}
\usepackage{graphicx}
\usepackage{color}

\begin{document}
\title{Interplay of paramagnetic, orbital and impurity effects on the
phase transition of a normal metal to superconducting state} 
\author{
M. Houzet and V. P. Mineev } 
\affiliation{Commissariat \`a l'Energie
Atomique, DSM/DRFMC/SPSMS, 38054 Grenoble, France} 
\date{June 29, 2006}

\begin{abstract}
We derive the generalized Ginzburg-Landau free
energy functional for conventional and unconventional singlet
superconductors in the presence of paramagnetic, orbital and impurity
effects. Within the mean field theory, we determine the criterion for
appearence of the non uniform (Fulde-Ferrell-Larkin-Ovchinnikov)
superconducting state, with vortex lattice structure and additional
modulation along the magnetic field. We also discuss the possible change of
the order of transition from normal to superconducting state. We find
that the superconducting phase diagram is very sensitive to
geometrical effects such as the nature of the order parameter and the
shape of the Fermi surface. In particular, we obtain the qualitative
phase diagrams for three-dimensional isotropic s-wave superconductors
and in quasi two-dimensional d-wave superconductors under magnetic
field perpendicular to the conducting layers.

In addition, we determine the criterion for instability toward non 
uniform superconducting state in s-wave superconductors in the dirty 
limit. 
\end{abstract}

\maketitle

\section{Introduction}

The relative importance of orbital and paramagnetic effects in
suppression of superconductivity is determined by the ratio of the 
orbital upper critical field \cite{Gor,Wert} $H_{c20} \approx
\Phi_{0} /2\pi\xi_{0a} \xi_{0b} $ and the paramagnetic limiting field
\cite{Clog, Chandr} $H_{p} = \Delta_{0} /\sqrt{2}\mu\approx
0.71\Delta_{0} /\mu$, called the Maki parameter \cite{Maki}
$\alpha_{M} =\sqrt{2}H_{c20} /H_{p} $.  Here $\Phi_{0} $ is the flux
quantum, $\xi_{0a} $ and $\xi_{0b} $ are the superconducting coherence
lengths in two mutually perpendicular and perpendicular to magnetic
field directions {\it a} and {\it b}, $\Delta_{0} $ is the
superconducting gap at zero temperature and $\mu=g\mu_{B} /2$ is the
electron magnetic moment.  Usually, the Maki parameter is of the 
order of the ratio of critical temperature to the Fermi energy
$\alpha_{M} \approx T_{c} /\varepsilon_{F} $. That demonstrates the
negligibly small influence of paramagnetic effects on
superconductivity.  However, in the case of small Fermi velocity (that
happens in materials with heavy electronic effective mass) or in the
layered metals under magnetic field parallel to the layers, the value
of the Maki parameter can be even larger than unity.  

The consideration of the magnetic field acting only on the electron spins,
corresponding to the limiting case of infinitely large Maki parameter,
leads to some peculiar effects.  First, the phase transition from
the normal metal to the superconducting state which is of the second
order in low field - high temperature region changes to the first 
order \cite{Sar,MaTs,Saint} at fields above $H^* \approx 1.06 T_{c} /\mu$ and
temperatures below $T^* \approx 0.56 T_{c} $.  Starting at this
critical point, the line of the first order transition is finished at
zero temperature and at the magnetic field equal to the
Chandrasekhar-Clogston limiting field $H_{p} $.  However, as was shown
by Fulde and Ferrell \cite{FF} and Larkin and Ovchinnikov \cite{LO},
even at larger field 
$H_{\text{FFLO}} \approx 0.755 \Delta_{0}/\mu \approx 1.07 H_{p}$,
the normal state is unstable with respect to the second order type
transition to the inhomogeneous cosine like gap modulated
superconducting state (FFLO state) with wave vector $q_{c} \approx 2.38\mu
H_{\text{FFLO}} /v_{F} $.  The recent calculations \cite{Bow} at zero
temperature have demonstrated that more complicated crystal structures
are more favorable than the simple plane wave. A first order type transition 
to the face-centered cube superconducting state was predicted to occur 
at the field larger than $H_{\text{FFLO}}$.  This conclusion is in
correspondence with the finite temperature investigations performed in
vicinity of critical point showing the appearence of FFLO
superconducting state below the critical temperature \cite{Buz,Mora}.

These results are changed a lot due to effects of orbital depairing 
and
impurities.

The role of the orbital effects was studied first at $T=0$
by Gruenberg and Gunther \cite{Gru} who have demonstrated 
that the FFLO state appears in pure metal (assuming that it is 
formed by means the second order transition) if the Maki parameter 
is larger than $1.8$.  

The influence of impurities in absence of the orbital effect was
investigated by Aslamazov \cite{Asl}.  He found that impurities do
not kill the FFLO state but decrease the field $H_{\text{FFLO}}$ of absolute
instability of the normal state for the FFLO formation such that, in
the dirty limit ($\tau T_{c}  \ll 1$), at zero temperature, $H_{\text{FFLO}}$
is lower than the field of the first order transition to homogeneous
superconducting state, $H_{p} $.  Physically, it does not yet abolish
a possibility of existence of inhomogeneous superconducting state 
because the actual phase transition from the normal state could be of the 
first order transition to FFLO state at some field $H>H_{p}$ .

The investigation of orbital effects near the critical point was
performed for isotropic three-dimensional pure metals by Houzet and Buzdin
\cite{Houz}.  It was found that, unlike the conclusions obtained in
the absence of orbital effect, for finite but large enough Maki
parameter, the FFLO modulated state arises from the normal state
starting from some temperature higher than the critical temperature.

All the studies cited above concerned the case of isotropic s-wave
superconductivity.  The theoretical interest to the FFLO state in
superconductors with d-pairing \cite{MakiWon, Sam,  Kun, Vor} was 
stimulated by the experimental identification of the pairing state
in several of the high-$T_{c} $ cuprate superconductors and heavy
fermionic materials.

The recently discovered heavy fermionic tetragonal compound $CeCoIn_{5}$ 
was established as $d_{x^{2} -y^{2} } $ superconductor similar to 
high-$T_{c}$ cuprates \cite{Mov, Iza}.  In this compound, the phase transition 
to the superconducting state becomes of the first order at low temperature - high
field region and possible formation of FFLO at lower temperatures
was reported for the magnetic field directed parallel \cite{Bian02, Mic,Bian03} 
as well as perpendicular \cite{Bian03} to the basal plane.

The first theoretical investigation of the phase diagram in the
tetragonal, doped-by-impurities superconductor with d-pairing under the
field parallel to c-axis was done in the Ref.\cite{Agt}. It was 
found that, in the absence of orbital effect, the change of the type of 
transition from the second to the first order occurs at some temperature which is
lower than the temperature of appearence of FFLO state.

The orbital effect in the same type of superconductor with quasi two-dimensional
spectrum was taken into account by Ikeda and Adachi \cite{Ik} and
the  different  phase diagram topology was established. That is,
in contrast with clean s-wave isotropic superconductor, the FFLO state 
arises from the normal state starting from some temperature lower than the critical
temperature.  This result was ascribed by the authors of Ref. \cite{Ik} 
to the nonpertubative treatment of the orbital effects incorporated there.

It seems, however, that, in absence of analytical calculations, it is 
dificult to recognize an inequivocal reason for this discrepancy. The main goal
of the present article is to make clear the influence of paramagnetic,
orbital and impurity effects on the phase transition of a normal metal
to superconducting state, including the FFLO state formation and type
of the phase transition.  With this purpose, we shall derive
Ginzburg-Landau functional for the conventional and
unconventional superconducting state with singlet pairing in the 
metal with arbitrary point symmetry and with arbitrary amount of point like
(s-wave scattering) impurities.  Then, for the cases of isotropic
metal with s-pairing and tetragonal superconductor with d-pairing
under magnetic field parallel to c-axis, the simple analytic criteria
of appearence of FFLO state and the type of normal-superconductor
phase transition shall be established. In particular, we shall
demonstrate which temperature of FFLO appearence or the critical point 
temperature is higher.

The structure of the article is as follows.  We begin with the general
expressions of Ginzburg-Landau functional for the superconducting 
state (in metal with the arbitrary concentration of impurities) transforming
according to identity and non-identity  representations of the crystal
point group symmetry.  The corresponding derivation of this functional
from microscopic theory valid at finite temperature in vicinity of
critical point is found in the Appendices.  Then, for the cases of
s-pairing and d-pairing, the criteria of FFLO state existence and
the first order type transition and their competition shall be
formulated.  In addition to these finite temperature calculations, the
critical field of dirty normal metal instabilty to FFLO state formation 
in presence of the orbital effect (generalization of the papers
by Gruenberg and Gunther \cite{Gru} and by Aslamazov \cite{Asl}) at
zero temperature is found.

\section{Free energy near critical point} \label{sec2}

The Ginzburg-Landau functional consists of the sum of the leading
terms in the expansion of the superconducting free energy in the order
parameter $\Delta$ and its gradients.

In the purely orbital limit, it contains terms proportional to 
$|\Delta |^2$, $|\Delta |^4$ and $|\nabla \Delta |^2$, with 
coefficients depending on the temperature and impurity concentration. 
Strictly speaking, it is only valid near the critical temperature 
$T_{c}$ of the second order transition from normal to superconducting 
state, when the coefficient $\alpha$ in front of $|\Delta |^2$ 
vanishes. Close to this point, the amplitude of the gap is indeed 
small and the magnetic length which determines the characteristic 
scale of the variation of $\Delta$ coincides with the thermal 
correlation length which diverges at $T_{c}$. This allows to retain only
the first term in gradient expansion.

In the purely paramagnetic limit, the coefficients in the functional 
also depend on the magnetic field. The equation $\alpha (T,H) =0$ 
then defines the transition line $H_0(T)$ from normal state to uniform 
superconducting state in the temperature-magnetic field phase 
diagram.  Along this line, the coefficient $\gamma$ in front of 
$|\nabla \Delta |^2$ happens to change its sign. This signals the instability
toward the modulated FFLO superconducting state.  In order to
establish the modulation wavelength in FFLO state, higher order terms
in the gradient expansion should also be included in the functional. 
One can restrict the free energy expansion to the term $|\nabla^2
\Delta |^2$ only in the vicinity of the triple point, where the FFLO
instability occurs and the typical FFLO modulation wavelength
diverges.  The coefficient $\beta$ in front of $|\Delta |^4$ also may
change its sign.  This signals a critical point when the type of the
transition into uniform superconducting state changes from second
order to first order as the temperature is lowered.  In such case,
the type of the transition into FFLO state will be determined by
the sign of the fourth order terms in $\Delta$ and of higher order in
the gradient expansion.  Again, close to the critical point, one can
consider the terms of the order $|\Delta |^2|\nabla \Delta |^2$ only
\cite{Buz}.  The peculiarity of the microscopic theory is that, in the
pure limit, $\gamma$ and $\beta$ change sign at the same place, with
coinciding triple and critical point: the tricritical point
$(T^*,H^*)$, with $H^*=H_0(T^*)$.  In the presence of impurities, however, 
the triple point and the critical point do not coincide any longer. 
For s-wave superconductors, the triple point occurs at lower temperature 
than the critical point, while for d-wave superconductors, the opposite
situation takes place \cite{Agt}.

The effect of the orbital field on the interplay between transition 
into conventional superconducting state (with vortex lattice in such 
case) and FFLO state (with FFLO modulation in the direction 
parallel to the vortex axes) was considered within this frame in 
s-wave superconductors and in pure limit only \cite{Houz}.  As it is
important that the magnetic length remains large compared to the
superconducting coherence length, a Ginzburg-Landau expansion is only
possible when the paramagnetic effect is much larger than the orbital
effect (large Maki parameter).  In the pure s-wave superconductor, it
was shown that the triple point was moved to higher temperatures 
\cite{Houz}. Therefore, impurities and orbital effect act in opposite 
directions in the s-wave case.

The goal of the two next Sections is to provide a frame to discuss 
the nature of the transition from normal to superconducting state in 
the presence of impurities and a small orbital effect in 
superconductors with arbitrary (even and one-component) order 
parameter.  
 
The free energy Ginzburg-Landau functional up to the fourth order 
terms of the order parameter and the fourth order terms in gradients 
for the isotropic s-pairing superconductors doped by impurities has
been derived first in the paper \cite{Ovc}.  With a purpose of
investigation of FFLO state   in clean s-wave superconductors
a similar result based on calculation using Eilenberger \cite{Eil} and
Larkin and Ovchinnikov \cite{LaOv} formalism has been derived also in 
the paper \cite{Houz}. The derivation for the dirty d-wave tetragonal
superconductor based on the direct calculation of the vertex parts
renormalization first introduced by Gor'kov \cite{Gor60} has been
accomplished in the paper \cite{Agt}, then including all orders in
gradients in the paper \cite{Ik}.  Our derivation made for reliability
by both Gor'kov \cite{Gor60} and Eilenberger \cite{Eil,LaOv} methods
(see Appendices A and B) is related to the case of doped-by-impurities
superconducting metal of arbitrary crystaline symmetry with an order
parameter
\begin{equation}
\Delta(\hat{\bf k},{\bf r})=\psi(\hat{\bf k})\Delta({\bf r})
\label{e1}
\end{equation}
transformed according to either identity, 
$\langle\psi(\hat{\bf k})\rangle \ne 0$, 
or non-identity, 
$\langle\psi(\hat{\bf k})\rangle = 0$,
one-dimensional representation of the point group symmetry of the
crystal \cite{Min}. Here and after, the angular brackets mean the 
averaging over the Fermi surface, $\psi(\hat{\bf k})$ are the functions 
of irreducible representations, $\langle|\psi(\hat{\bf k})|^{2} \rangle= 1$.  
The generalization for the multidimensional superconducting states can be 
easily considered.

The derivation for the case of superconducting state with an order
parameter transforming according to general identity representation 
leads to quite cumbersome expression for the free energy. 
We shall consider the simplest example of identity representation with
$\psi(\hat{\bf k})=1$, or s-wave pairing superconductivity, where the 
free energy functional is
\begin{eqnarray}
&&F =\int d^{3} r\left\{\alpha|\Delta|^{2}  + \pi
N_{0} \Delta^{*} \left[\frac{K_{21} }{4}\langle ({\bf v}{\bf
D})^{2} \rangle \right.\right.\nonumber \\
&&\left.-\left.\frac{K_{23} }{16}\langle({\bf v}{\bf D})^{4} \rangle-
\frac{K_{33} }{32\tau  }\langle({\bf v}{\bf
D})^{2} \rangle^{2} \right]\Delta \right.\nonumber \\
&&\left.+ \pi N_{0} \frac{K_{30} }{4}
|\Delta|^{4}  \right.\nonumber \\
&&\left.+\pi N_{0} \left(
-\frac{K_{41} }{2}+\frac{K_{42} }{16\tau}\right)|\Delta|^{2} 
\langle
({\bf v}{\bf D}\Delta)^{*}  ({\bf v}{\bf
D}\Delta)\rangle\right.\nonumber \\
&&\left.+\pi N_{0} \frac{K_{41} }{16}\left[(\Delta^{*} )^{2} 
\langle({\bf v}{\bf D}\Delta)^{2} \rangle+ c.c. \right] \right\}.
\label{e2}
\end{eqnarray}
Here, 
\begin{equation}
\alpha=N_{0} \Re\left[\Psi\left(\frac{1}{2}- \frac{i\mu
H}{2\pi T}\right)-\Psi\left(\frac{1}{2}- \frac{i\mu
H_{0} }{2\pi T}\right)\right]
\label{e3}
\end{equation}
and $H_{0} =H_{0} (T)$ is the critical field in homogeneous 
superconductor
determined by the equation
\begin{equation}
\ln \frac{T_{c} }{T}=\Re\left[\Psi\left(\frac{1}{2}- \frac{i\mu
H_{0} }{2\pi T}\right)-\Psi\left(\frac{1}{2}\right) \right].
\label{e4}
\end{equation}
The coefficients
\begin{equation}
K_{nm} =2T\Re{\sum^{\infty} }_{\nu=o} \frac{1}{(\omega_{\nu} -i\mu
H)^{n}  (\tilde\omega_{\nu} -i\mu H)^{m} },
\label{e6}
\end{equation}
and $\omega_{\nu} =\pi T(2\nu +1)$ are Matsubara frequencies,
$$
\tilde\omega_{\nu} =\omega_{\nu} +\frac{\text{sign} ~\omega_{\nu} }{2\tau},
$$
${\bf D}=-i\nabla+(2\pi /\Phi_{0} ){\bf A}$, ${\bf v}(\hat{\bf k})$ is
the Fermi velocity, $N_0$ is the density of states at the Fermi level.
We put through the article $\hbar=1$.

Near the critical temperature $T_{c} $, both fields $H_{0} (T)$ and
the upper critical field tend to zero and the latter should be 
determined from the linear Ginzburg-Landau equation giving well known Gor'kov 
result \cite{Gor60} with small correction due to paramagnetic 
effect. On the contrary, near the tricritical point, at large enough Maki 
parameter the upper critical field is close to $H_{0} (T)$ such that one can
write
\begin{eqnarray}
   \alpha &=& \alpha_{0} (H-H_{0}) \nonumber \\
    &=&
\frac{N_{0} \mu (H-H_{0} )}{2\pi T}
\Im\Psi^{\prime} \left(\frac{1}{2}- \frac{i\mu H_{0} }{2\pi T}\right)
\label{e5}
\end{eqnarray}
and put $H=H_{0} $ in all other terms of the functional.

The free energy functional for non-identity representation is
\begin{eqnarray}
&&
F = \int  d^{3} r
\left\{
	\tilde\alpha|\Delta|^{2} 
\right.
\nonumber \\
&&
+
\left.
	\pi N_{0} \Delta^{*} 
	\left[
		\frac{K_{03} }{4}
		\langle |\psi(\hat{\bf k})|^{2} ({\bf v}{\bf D})^{2} \rangle        
\right. 
\right.
\nonumber \\
&&
\left.
	-
	\left.
		\frac{K_{05} }{16}
		\langle|\psi(\hat{\bf k})|^{2} ({\bf v}{\bf D})^{4} \rangle 
	\right.
\right.
\nonumber \\
&&
\left.
- 
\frac{K_{15} }{32\tau}
\langle\psi(\hat{\bf k})^*({\bf v}{\bf D})^{2} \rangle
\langle\psi(\hat{\bf k})({\bf v}{\bf D})^{2} \rangle  
\right]
\Delta 
\nonumber \\
&&\left.+ \pi N_{0} \left(\frac{K_{03} }{4} \langle|\psi(\hat{\bf
k})|^{4} \rangle-\frac{K_{04} }{8\tau}\right)|\Delta|^{4} 
\right.\nonumber \\
&&\left.
-\pi N_{0} \frac{K_{05} }{2}|\Delta|^{2}  \langle|\psi(\hat{\bf
k})|^{4} ({\bf v}{\bf D}\Delta)^{*}  ({\bf v}{\bf
D}\Delta)\rangle\right.\nonumber \\
&&\left.+\pi N_{0} \frac{5 K_{06} }{16\tau}|\Delta|^{2}  
\langle|\psi(\hat{\bf
k})|^{2} ({\bf v}{\bf D}\Delta)^{*}  ({\bf v}{\bf
D}\Delta)\rangle\right.\nonumber \\
&&\left.+\pi N_{0} \frac{K_{05} }{16}\left[(\Delta^{*} )^{2}  
\langle|\psi(\hat{\bf
k})|^{4} ({\bf v}{\bf D}\Delta)^{2} \rangle+ c.c. 
\right]\right.\nonumber \\ 
&&\left.-\pi N_{0} \frac{K_{06} }{16\tau}\left[(\Delta^{*} )^{2}  
\langle|\psi(\hat{\bf
k})|^{2} ({\bf v}{\bf D}\Delta)^{2} \rangle+ c.c. \right]
\right\}.
\label{e7}
\end{eqnarray}
Here, 
\begin{eqnarray}
&&\tilde \alpha=N_{0} \Re\left[\Psi\left(\frac{1}{2}- \frac{i\mu 
H}{2\pi
T}+\frac{1}{4\pi\tau T}\right)\right.\nonumber \\
&&\left.-\Psi\left(\frac{1}{2}- \frac{i\mu \tilde H_{0} }{2\pi
T}+\frac{1}{4\pi\tau T}\right)\right]
\label{e8}
\end{eqnarray}
and $\tilde H_{0} =\tilde H_{0} (T)$ is the critical field in 
homogeneous
superconductor determined by the equation
\begin{equation}
\ln \frac{T_{c} }{T}=\Re\left[\Psi\left(\frac{1}{2}- \frac{i\mu
\tilde H_{0} }{2\pi T}+\frac{1}{4\pi\tau 
T}\right)-\Psi\left(\frac{1}{2}\right) \right].
\label{e9}
\end{equation}
Near the critical temperature $T_{c} $, both fields 
$\tilde H_{0} (T)$ and the upper critical field tend to zero and 
the latter should be determined from the linear Ginzburg-Landau equation.  
Near the tricritical point, at large enough Maki parameter, the upper critical
field is close to $\tilde H_{0} (T)$ such that one can write
\begin{eqnarray}
\tilde \alpha&=&\tilde \alpha_{0} (H-\tilde H_{0} )\nonumber \\
&=&\frac{N_{0} \mu
(H-\tilde H_{0} )}{2\pi T} \Im\Psi^{\prime} \left(\frac{1}{2}- 
\frac{i\mu
\tilde H_{0} }{2\pi T}+\frac{1}{4\pi\tau T}\right)
\label{e10}
\end{eqnarray}
and put $H=\tilde H_{0} $ in all other terms of the functional.

In addition to the superconducting energy (\ref{e2}) or (\ref{e7}), 
one should in principle also include the magnetic energy
\begin{equation}
F_m = \frac{1}{8\pi}
\int d^3r ({\bf H-H_{\text{ext}}})^2,
\end{equation}
where $\bf H_{\text{ext}}$ is the external field, in the total free 
energy functional. In the following, we will negelect the 
contribution of such term by assuming that the screening current are 
not important (high-$\kappa$ limit) and $\bf H=H_{\text{ext}}$.

\section{Criteria for the appearence of FFLO state and the first order
transition }

\subsection{Identity representation}

With the purpose to derive simple analytic criteria for the appearence
of FFLO state and the change of the second order normal metal -
superconductor transition to the first order one, let us make the
angular averaging in the expression (\ref{e2}) for the free energy in 
the case of pure s-pairing ($\psi(\hat{\bf k})=1$) in a metal with spherical 
Fermi suface:
\begin{eqnarray}
&&F =\int d^{3} r\left\{\alpha_{0} (H-H_{0} )|\Delta|^{2}  + 
\frac{\pi N_{0}v^{2} K_{21} }{12} \Delta^{*} ({\bf D})^{2} \Delta
\right.\nonumber \\
&&\left.-\frac{\pi N_{0}v^{4} K_{23} }{80}\Delta^{*} \left ( ({\bf
D}^{2} )^{2} +\frac{1}{\lambda^{4} } \right)\Delta\right.\nonumber\\
&&\left.- \frac{\pi N_{0} v^{4} K_{33} }{288\tau  } \Delta^{*} 
({\bf D}^{2} )^{2} \Delta + \frac{\pi N_{0} K_{30} }{4} |\Delta|^{4} 
\right.\nonumber\\
&&\left.+\pi N_{0} v^{2} \left(
-\frac{K_{41} }{6}+\frac{K_{42} }{48\tau}\right)|\Delta|^{2} 
|{\bf D}\Delta|^2\right.\nonumber \\
&&\left.+\frac{\pi N_{0} v^{2} K_{41} }{48} \left[(\Delta^{*} )^{2} 
({\bf D}\Delta)^{2} + c.c. \right] \right\},
\label{e11}
\end{eqnarray}
where $v$ is the modulus of Fermi velocity and the term
$\lambda^{-4} =(2eH/c)^{2} $ originates from noncommutativity of the
operators $D_{x} $ and $D_{y} $.  This value serves as the measure of
the orbital effect such that the orbital effects free situation
corresponds to the limit $\lambda \to \infty$.

Let us choose the magnetic field direction along the z-axis ${\bf
A}=(0,Hx,0))$. So, for the Abrikosov lattice ground state
$\Delta=\varphi_{0} (x,y)f(z)$ which is the linear combination of 
Landau wave
fuctions with $n=0$ multiplied by exponentially 
$f_{\text{exp}}=\exp(iqz)$ or sinusoidally
$f_{\text{sin}}=\sqrt{2}\sin qz$ modulated function along z-direction, 
one can substitute
\begin{equation}
{\bf D}^{2} \Delta= ({\bf D}_{\perp} ^{2} +D_{z} ^{2} )\Delta=
\left(\frac{1}{\lambda^{2} }+q^{2} \right)\Delta.
\label{e12}
\end{equation}
Making use of the properties: 
\begin{equation} \label{eq:prop}
({\bf D}_{\perp} \Delta)^{2} =0,~~~\int
d^{3} r|\Delta|^{2} |{\bf 
D}_{\perp} \Delta|^{2} =\frac{1}{2\lambda^2}\int
d^{3} r|\Delta|^{4} ,
\end{equation} 
we come to the free energy in the following form
\begin{equation} \label{e13}
F=
\int dxdy
\left\{
\alpha_0\left(H-H(q)\right)
|\varphi_0|^2
+
\pi N_{0}  B
|\varphi_0|^4
\right\},
\end{equation}
where
\begin{eqnarray} \label{e14}
&&H(q)=H_{0} -\frac{\pi N_{0}}{\alpha_{0} } 
\left[\frac{v^{2} K_{21} }{12}
\left(\frac{1}{\lambda^{2} }+q^{2} \right) \right.
\\
&&\left.-\frac{v^{4} K_{23} }{80}
\left(\left(\frac{1}{\lambda^{2} }+q^{2} \right)^{2} 
+\frac{1}{\lambda^{4} } \right)\right.
\left.- \frac{v^{4} K_{33} }{288\tau }
\left(\frac{1}{\lambda^{2} }+q^{2} \right)^{2}  \right],
\nonumber
\end{eqnarray}
and the coefficient $B$ is given by
\begin{equation}
B_0=
\frac{K_{30} }{4}
-\frac{v^{2}}{2\lambda^2} \left(
\frac{K_{41} }{6}-\frac{K_{42} }{48\tau}\right)
\end{equation}
in the conventional superconducting vortex-lattice state, 
\begin{equation}
B_{\text{exp}}=
B_0
-v^{2} q^2 
\left(
\frac{K_{41} }{8}-\frac{K_{42} }{48\tau}\right)
\end{equation}
in the exponentially modulated FFLO phase, and
\begin{equation}
B_{\text{sin}}=
\frac{3B_0 }{2}
-\frac{q^2 v^{2}}{2} \left(
\frac{5K_{41} }{24}-\frac{K_{42} }{48\tau}\right)
\end{equation}
in the sinusoidally modulated FFLO phase.

The critical field value $H_{c} $ is found by taking the maximum of 
$H=H(q)$ as the function of $q$.  The usual superconducting state appears at 
$q=0$, while the FFLO state is formed when the
maximum of $H$ is reached at finite $q=q_{0} $, where
\begin{equation}
q_{0} ^{2} =-\frac{1}{\lambda^{2} }+\frac{\frac{K_{21} }{3v^{2} }}
{\frac{K_{23} }{10}+\frac{K_{33} }{36\tau}}.
\label{e15}
\end{equation}
The FFLO state appears when the coefficient at $q^{2} $ in the
square brackets of Eq.~(\ref{e14}) changes the sign from positive to
negative and it exists at
\begin{equation}
K_{21} ~<~\frac{3v^{2} }{\lambda^{2} }
\left(\frac{K_{23} }{10}+\frac{K_{33} }{36\tau}\right).
\label{e16}
\end{equation}

Let us now examine the question of the transition type.  It is
determined by the sign of the coefficient at $|\varphi_0|^{4} $ in the
expression (\ref{e13}) for the free energy.  Hence, the first order
transition occurs at $B_0<0$ for the transition into
usual superconducting state.  It occurs at $B_{\text{exp}}(q_0)<0$ for
transition into FFLO state with exponential modulation or
$B_{\text{sin}}(q_0)<0$ for transition into FFLO state with sinusoidal
modulation.

To see explicitly the role of orbital effects and the impurities in 
the formation of FFLO state and the change of the transition type, let us
look on them separetely.  

\begin{figure}[tbp]
\includegraphics[scale=1]{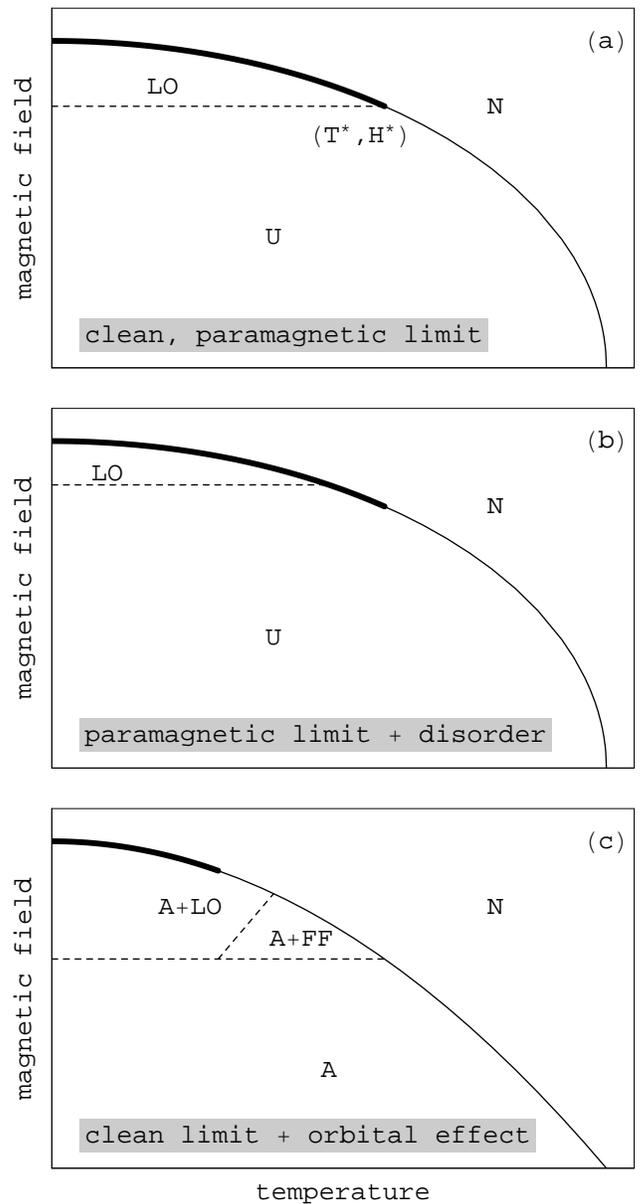}
\caption{
Qualitative superconducting phase diagram for three-dimensional 
s-wave superconductor in presence of strong paramagnetic effect.  The 
clean, purely paramagnetic case is shown on (a), the purely 
paramagnetic case with disorder is shown on (b), 
and the clean case, with some orbital effect is 
shown on (c). Possible phases are the normal state (N), the 
conventional superconducting state: uniform state (U) in the absence 
or orbital effect, Abrikosov vortex lattice state (A) in presence of 
orbital effect, and the FFLO modulated state with exponential 
modulation (FF) or sinusoidal modulation (LO) along the applied 
field, eventually
with Abrikosov vortex lattice (A+FF or A+LO) if orbital effect is 
present. Thin lines correspond 
to second order transitions, thick lines correspond to first order 
transitions, dashed lines correspond to transitions between different 
superconducting states and they have not been calculated in the 
present work. 
}
\label{fig1}
\end{figure}

In the clean, paramagnetic limit ($\lambda,\tau=\infty$), 
Eq.~(\ref{e13}) coincides with the free energy derived in Ref.~\cite{Buz}. 
There, the inequality $K_{30}<0$ was obtained as the condition both for 
the change of the transition type from normal to uniform superconducting state 
and for the FFLO state formation. In the vicinity of the tricritical point, 
keeping $H=H_0(T)$ in $K_{30}$, one finds that $K_{30}$ changes its sign as a function 
of the temperature when $T=T^*$. Therefore, $K_{30}\propto (T-T^*)$ has the meaning 
of an effective temperature close to this point. Further study reveals that, 
while $B_{\text{exp}}(q_0)=-K_{30}/6$ remains positive at
$T<T_t$, $B_{\text{sin}}(q_0)=K_{30}/36$ becomes negative. This means 
that the first order transition from sinusoidally modulated FFLO state is 
favored at $T<T^*$. The qualitative superconducting phase diagram which results 
from this study is shown on Fig.~\ref{fig1}(a).

In presence of impurities but neglecting the orbital effect
($\lambda \to \infty $), we obtain from Eq.~(\ref{e16}) inequality $K_{21} <0$ 
as condition  of FFLO formation.  It is easy to check that at the temperature 
determined by equation $K_{21} =0$, where the coefficient $K_{21} $ changes the 
sign and, hence, the finite $q$ modulation appears, the coefficient $K_{30}$
is already negative. Therefore, $B_0$ is negative and the normal metal transforms to
homogeneous superconducting state by means of the first order
transition.  The impurities shift FFLO state to lower temperatures
leaving unchanged the temperature of change of type of transition. 
The qualitative phase diagram in this limit is shown on
Fig.~\ref{fig1}(b).

On the other hand, taking into account only orbital effects,
that is in the completely pure case ($\tau \to \infty$), one can rewrite
the condition (\ref{e16}) of FFLO appearence as
\begin{equation}
K_{30} ~<~\frac{3}{10}\frac{v^{2} K_{50} }{\lambda^{2} }.
\label{e18}
\end{equation}
In pure case and at $\lambda \to \infty$, the FFLO state appears
exactly when the coefficient $K_{30} $ changes its sign.  Whereas in
presence of orbital effect, the FFLO state appears at slightly 
negative $K_{30}$ determined by negative value of $K_{50} $.
Moreover, the condition of the first order transition into usual 
superconducting state, $B_0<0$, is rewritten as
\begin{equation}
K_{30} ~<~\frac{1}{3}\frac{v^{2} K_{50} }{\lambda^{2} }
\label{e19}
\end{equation}
The comparison of these two inequalities makes clear that, due to the
orbital effect, the change of the type of transition always appears at
lower temperature than the FFLO state formation.  This conclusion is
in correspondence with the results of the paper \cite{Houz}  
where the qualitative phase diagram shown on Fig.~\ref{fig1}(c) was first 
proposed.

Thus, we find that impurities and orbital effect act in opposite 
directions regarding the shift of the temperature below which FFLO 
state will appear. In the following, we study the interplay between 
low impurities and orbital effect ($1/\lambda,1/\tau\rightarrow 0$) 
on FFLO state formation. In this limit, the change of the transition 
type at the normal/conventional superconducting vortex lattice 
transition, determined by $B_0<0$, is still given by 
Eq.~(\ref{e19}). The temperature below which transition into FFLO 
state occurs is determined by Eq.~(\ref{e16}). In leading order in 
$1/\lambda^2$ and $1/\tau$, this equation yields:
\begin{equation}
K_{30}
<
\frac{1}{2}\frac{K_{40}}{\tau}
+\frac{3}{10}\frac{v^2 K_{50}}{\lambda^2},
\label{e19b}
\end{equation}
where we made use of the property :
$$
K_{nm}\simeq K_{n+m,0}-\frac{m}{2\tau}K_{n+m+1,0}.
$$
In Eqs.~(\ref{e19}), (\ref{e19b}), we recall that $K_{30}\propto 
(T-T^*)$ has the meaning of an effective temperature, while $K_{40}$ 
and $K_{50}$ have to be evaluated at the tricritical point 
$(T^*,H^*)$, where they take negative values. By comparing 
Eqs.~(\ref{e19}),(\ref{e19b}), we find that the FFLO state appears at 
temperatures higher than the critical temperature when the impurity 
concentration remains low enough:
\begin{equation} \label{e19bb}
\frac{1}{\tau}<\frac{v^2}{15 \lambda^2}\frac{K_{50}}{K_{40}}.
\end{equation}

When Eq.~(\ref{e19bb}) is obeyed, the free energy (\ref{e13}) also 
allows to discuss the structure of the FFLO state which is realized 
at the second order normal/FFLO transition. When 
$B_{\text{exp}}(q_0)<B_{\text{sin}}(q_0)$ (and both of them 
positive), with $q_0$ given by Eq.~(\ref{e15}), the FFLO state 
with exponential modulation is energetically favored. In the limit of 
low impurity and orbital effect that we consider, this inequality 
corresponds to:
\begin{equation}
K_{30}
>
\frac{5}{28}\frac{K_{40}}{\tau}
+
\frac{9}{28}\frac{v^2 K_{50}}{\lambda^2}.
\label{e19c}
\end{equation}
The FFLO state with sinusoidal modulation is favored when 
$B_{\text{sin}}(q_0)<B_{\text{exp}}(q_0)$. The transition into this 
state becomes of the first order when $B_{\text{sin}}(q_0)<0$, that 
is:
\begin{equation}
K_{30}
<
-\frac{25}{4}\frac{K_{40}}{\tau}
+\frac{3}{4}\frac{v^2K_{50}}{\lambda^2}.
\label{e19d}
\end{equation}

\begin{figure}[tbp]
\includegraphics[scale=1]{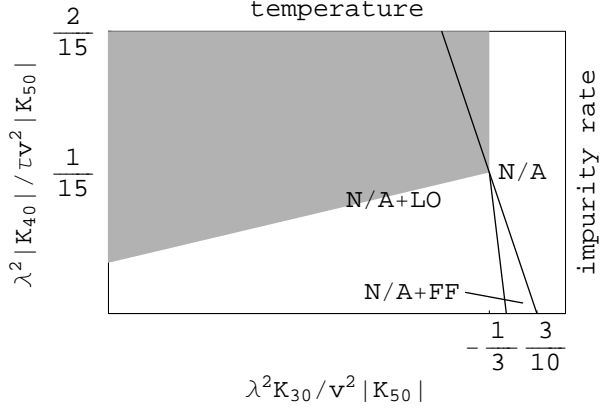}
\caption{
This figure shows the nature of the superconducting state which is 
realized just below the upper critical line, as well as the type of the 
transition, for three-dimensional s-wave superconductors with spherical 
Fermi surface, at large paramagnetic, small orbital and impurity 
effects, and for temperatures close to the tricritical temperature 
$T^*$. $\lambda^2K_{30}/v^2|K_{50}| \propto (T-T^*)$ is the effective 
temperature, $\lambda^2|K_{40}|/\tau v^2|K_{50}| \propto \tau^{-1}$ is 
the ratio between orbital and impurity effects. N/A indicates the 
place  where the transition is from normal state into conventional Abrikosov 
superconducting vortex lattice state, N/A+FF indicates the place where the 
transition is from normal state into FFLO state with vortex structure in the plane 
parpendicular to the magnetic field and exponential modulation along the field, 
N/A+LO indicates the place where the transition is into FFLO state with 
vortex structure in the plane perpendicular to the magnetic field and 
sinusoidal modulation along the field; in the white (gray) region, 
the transition into superconducting state is of the second (first) order.
}
\label{fig2}
\end{figure}

The above discussion is summarized on Fig.~\ref{fig2}. It represents 
the nature and type of the transition from normal to the different 
superconducting states as temperature is lowered, for a given ratio 
between orbital and impurity effects.

When the transition is of the second order, minimization of the free 
energy (\ref{e13}) on $\varphi_0$ yields that the vortex lattice is 
triangular.\cite{Abrikosov}

In the Ref.\cite{BuzBris}, it was predicted that a very large Maki 
parameter may favor not only FFLO modulation, but also an order 
parameter formed of higher-level Landau functions: $\Delta\sim 
\varphi_n(x,y) f(z)$, with $n>0$. In Eqs.~(\ref{eq:prop}),(\ref{e14}) 
the expression $(\lambda^{-2}+q^2)$ should be then replaced by 
$((2n+1)/\lambda^2+q^2)$, and the fourth order term $B$ in the free 
energy (\ref{e12}) should also be calculated accordingly. 
In particular, the condition to maximize the critical field 
(\ref{e14}) at the second order transition from normal to vortex lattice 
state with Landau functions of level $n \geq 1$ is:
\begin{equation}
K_{30}
<
\frac{1}{2}\frac{K_{40}}{\tau}
+\frac{3(2n+1)}{10}\frac{v^2 K_{50}}{\lambda^2}.
\end{equation}
This equation is obtained by requiring that $q_0^2>0$ in 
Eq.~(\ref{e15}) after substituting
$1/\lambda^2$ by $(2n+1)/\lambda^2$.
One can note that, when such inequality is obeyed, the transition from normal 
to vortex lattice state (with $n=0$) has already turned from second to first 
order, both in the "low impurity" case in the presence
of sinusoidal modulation (because Eq.~(\ref{e19d}) is already 
obeyed), and in the "high impurity" case
(because Eq.~(\ref{e19}) is already obeyed).

Therefore, we get now the qualitative picture of the superconducting phase 
diagram in three-dimensional s-wave superconductors with strong paramagnetic effect.
At large impurity concentration, the transition from 
normal to usual superconducting state becomes of the first order at 
low temperatures, while FFLO state may exist at even more lower
temperatures either as stable or as metastable state. 
On the other hand, at low impurity concentration, while temperature is lowered, 
the phase diagram shows the second order transition from normal to usual
supercondutcing state, then to exponential FFLO state, then to
sinusoidal FFLO state, and finally the change of the transition order
into such state.  These conclusions are summarized in the phase
diagrams shown in Fig.~\ref{fig1} and \ref{fig2}.

\subsection{Non-identity representation}

As an example of similar calculations for nonconventional
superconductivity we consider the d-wave superconducting state
$\psi(\hat{\bf k})\propto k_{x} ^{2} - k_{y} ^{2} $  in
tetragonal crystal under magnetic field along c-axis ($\hat
z$-direction).  One can rewrite first the Eq.~  (\ref{e7}) in the
following form
\begin{eqnarray}
&&F =\int
d^{3} r\left\{\tilde\alpha|\Delta|^{2} \right.\nonumber \\
&&+
\left.\pi N_{0} \Delta^{*} \left[\frac{K_{03} }{4}(
\langle|\psi|^{2} v_{\perp} ^{2} /2\rangle {\bf D}_{\perp} ^{2} +
\langle|\psi|^{2}  v_{z} ^{2} \rangle 
D_{z} ^{2} )\right.\right.\nonumber \\
&&\left.\left.-\frac{K_{05} }{16}(\langle|\psi|^{2} ({\bf v}{\bf
D}_{\perp} )^{4} \rangle+3\langle|\psi|^{2} v_{\perp} ^{2} v_{z} ^{2} \rangle 
{\bf D}_{\perp} ^{2}  D_{z} ^{2} \right.\right.\nonumber \\
&&\left.\left.+\langle|\psi|^{2} v_{z} ^{4} \rangle  
D_{z} ^{4} )\right.\right.\nonumber \\
&&\left.\left.- \frac{K_{15} }{32\tau}|\langle\psi\hat 
v_{\perp} ^{2} /2\rangle{\bf
D}_{\perp} ^{2} +\langle\psi\hat v_{z} ^{2} \rangle
D_{z} ^{2} |^{2} \right]\Delta \right.\nonumber \\
&&\left.+ \pi N_{0} \left(\frac{K_{03} }{4}
\langle|\psi|^{4} \rangle-\frac{K_{04} }{8\tau}\right)|\Delta|^{4} 
\right.\nonumber \\
&&\left.
-\pi N_{0} \frac{K_{05} }{2}|\Delta|^{2} 
\left[\langle|\psi|^{4} v_{\perp} ^{2} /2\rangle({\bf
D}_{\perp} \Delta)^{*}  {\bf D}_{\perp} \Delta\right.\right.\nonumber 
\\
&&\left.\left.+\langle|\psi|^{4} v_{z} ^{2} \rangle( 
D_{z} \Delta)^{*}  
D_{z} \Delta\right]\right.\nonumber \\
&&\left.+\pi N_{0} \frac{5 K_{06} }{16\tau}|\Delta|^{2} 
\left[\langle|\psi|^{2} v_{\perp} ^{2} /2\rangle({\bf
D}_{\perp} \Delta)^{*}  {\bf D}_{\perp} \Delta\right.\right.\nonumber 
\\
&&\left.\left.+\langle|\psi|^{2} v_{z} ^{2} \rangle(D_{z} \Delta)^{*}  
D_{z} \Delta\right]\right.\nonumber \\
&&\left.+\pi N_{0} \frac{K_{05} }{16}\left[(\Delta^{*} )^{2} 
(\langle|\psi|^{4}  v_{\perp}^{2} /2\rangle({\bf
D}_{\perp} \Delta)^{2} \right.\right.\nonumber \\
&&\left.\left.+(\Delta^{*} )^{2} \langle|\psi|^{4}  v_{z}^{2}\rangle(
D_{z} \Delta)^{2} )+ c.c. \right] \right.\nonumber \\
&&\left.-\pi N_{0} \frac{K_{06} }{16\tau}\left[(\Delta^{*} )^{2} 
(\langle|\psi|^{2}  v_{\perp}^{2} /2\rangle({\bf
D}_{\perp} \Delta)^{2} \right.\right.\nonumber \\
&&\left.\left.+(\Delta^{*} )^{2} \langle|\psi|^{2}  v_{z}^{2}\rangle(
D_{z} \Delta)^{2} )+ c.c. \right] \right\}.
\label{e20}
\end{eqnarray}

From this step, unlike to the conventional superconductivity, the
continuation of calculation for d-pairing in closed analytical form is
not possible.  The point is that the average 
$\langle|\psi|^{2} ({\bf v}{\bf D}_{\perp} )^{4} 
\rangle=\langle|\psi|^{2} (D^{+} v^{-} +D^{-} v^{+} )^{4} \rangle$ 
contains the terms
$\langle|\psi|^{2} (D^{\pm} v^{\mp} )^{4} \rangle$ which are not equal
to zero in tetragonal crystal.  Here 
$D^{\pm} =(D_{x} \pm iD_{y} )/\sqrt 2$, $v^{\pm} =(v_{x} \pm iv_{y} )/\sqrt 2$ 
and $v_{\perp} ^{2} =v_{x} ^{2} +v_{y} ^{2} $.  Hence, unlike to 
conventional superconductivity, the Abrikosov lattice ground state in tetragonal
superconductor with d-pairing \cite{Vav} is the linear combination of
functions consisting of infinite series of Landau wave functions
$\varphi_{n} (x,y)$ with n=0, 4, 8, 12\ldots multiplied by
exponentially or sinusoidally modulated function $f(z)$ along
z-direction
\begin{equation}
\Delta=f(z)(A_{0} \varphi_{0} +A_{4} \varphi_{4} 
+A_{8} \varphi_{8} \ldots.)
\label{e21}
\end{equation}
Fortunately, in the limit of large Maki parameters we are interested 
in, one can work with cut-off series of the form similar to s-wave pairing
\begin{equation}
\Delta\approx f(z)\varphi_{0} 
\label{e22}
\end{equation}
and also neglect the terms like 
$\langle|\psi|^{2} (D^{\pm} v^{\mp} )^{4} \rangle$ 
in the Hamiltonian (the proof of this property is found in
Appendix C).  So, we shall use the substitution
\begin{eqnarray}
&&\langle|\psi|^{2} ({\bf v}{\bf 
D}_{\perp} )^{4} \rangle\Delta\nonumber \\
&&\Rightarrow~~
\langle|\psi|^{2} (v^{-} v^{+} )^{2} \rangle(D^{-} D^{+} D^{-} D^{+} 
+D^{-} D^{-} D^{+} D^{+} )\Delta\nonumber \\
&&=3\langle|\psi|^{2} (v^{-} v^{+} )^{2} \rangle\lambda^{-4} \Delta 
\label{e23}
\end{eqnarray}
Thus, the further calculations have the sence of
variational treatment.

Similar to the case of conventional superconductivity, we now obtain
from Eq.~ (\ref{e20})
\begin{equation} \label{e21}
F=
\int dxdy
\left\{
\tilde{\alpha}_0\left(H-\tilde{H}(q)\right)
|\varphi_0|^2
+
\pi N(0) \tilde{B}
|\varphi_0|^4
\right\},
\end{equation}
where
\begin{eqnarray}
&&
\tilde{H}(q)
=
\tilde{H}_{0}
-
\frac{\pi N_{0}}{\tilde\alpha_{0}}
\left[
	\frac{K_{03}}{4}
	(
		\langle |\psi|^2v_\perp^2\rangle
		\frac{1}{2\lambda^2}
		+
		\langle |\psi|^2v_z^2\rangle
		q^{2}
	)
\right.
\nonumber\\
&&
\left.
	-
	\frac{K_{05}}{16}
		(
			\frac{3}{4 \lambda^4}
			\langle |\psi|^2v_\perp^4\rangle
			+
		    \frac{3 q^2}{ \lambda^2}
		    \langle |\psi|^2v_\perp^2v_z^2\rangle
		    +
		    \langle |\psi|^2v_z^4\rangle q^4
		 )
\right.	 
\nonumber\\
&&
\left.
	-
	\frac{K_{15}}{32\tau}
		|
			\langle \psi v_\perp^2\rangle
			\frac{1}{2\lambda^2}
			+
			\langle \psi v_z^2\rangle
			q^{2}
		 |^2
\right]	 
\label{e25}
\end{eqnarray}
and 
\begin{equation}
\tilde B_0
=
\frac{K_{03}}{4}
\langle|\psi|^4\rangle
-\frac{K_{04}}{8\tau}
-\frac{K_{05}}{8\lambda^2}
\langle |\psi|^4 v_\perp^2\rangle
+\frac{5K_{06}}{64\tau\lambda^2}
\langle |\psi|^2 v_\perp^2\rangle
\end{equation}
for the conventional superconducting vortex lattice state,
\begin{equation} \label{eBexpd}
\tilde B_{\text{exp}}(q)
= 
\tilde B_0
+
q^2
\left(
	-\frac{3 K_{05}}{8} \langle |\psi|^4 v_z^2\rangle
	+\frac{3 K_{06}}{16\tau} \langle |\psi|^2 v_z^2\rangle
\right)
\end{equation}
for the exponentially modulated FFLO state, and
\begin{equation}\label{eBsind}
\tilde B_{\text{sin}}(q)
= 
\frac{3\tilde B_0}{2}
+
\frac{q^2}{2}
\left(
	-\frac{5 K_{05}}{8} \langle |\psi|^4 v_z^2\rangle
	+\frac{7 K_{06}}{16\tau} \langle |\psi|^2 v_z^2\rangle
\right)
\end{equation}
for the sinusoidally modulated FFLO state.

\begin{figure}[tbp]
\includegraphics[scale=1]{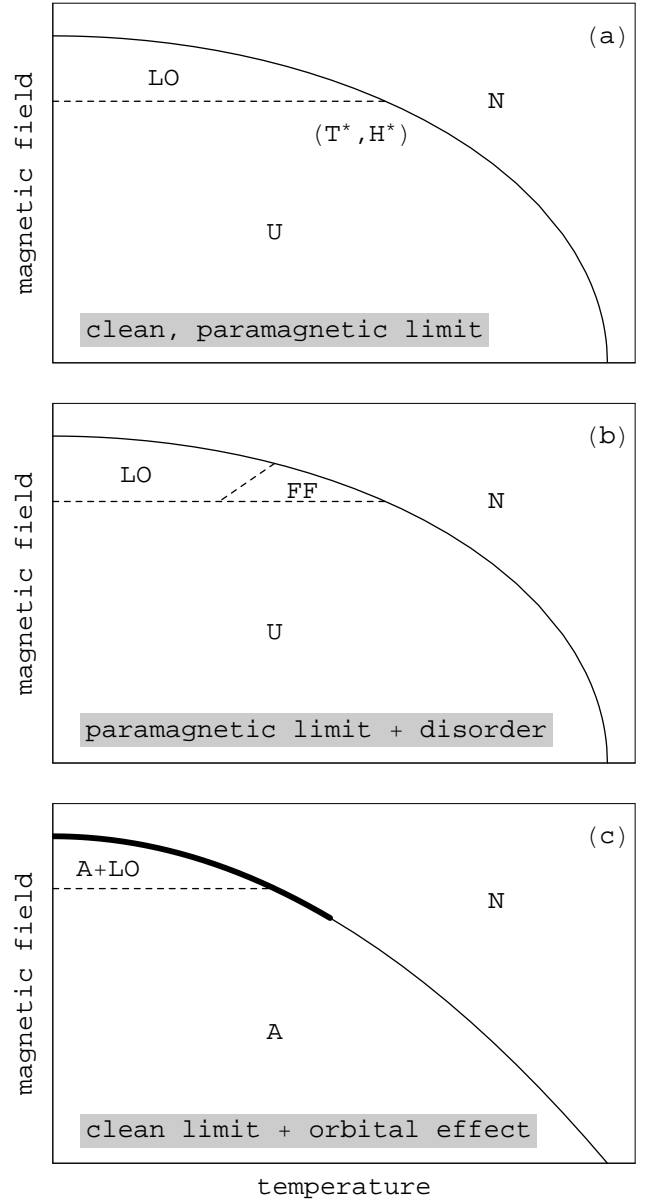}
\caption{
Qualitative superconducting phase diagram for quasi two-dimensional 
d-wave superconductor in presence of strong paramagnetic effect.  
(see also legend of Fig.~\ref{fig1}). 
}
\label{fig3}
\end{figure}

The study is now similar to the previous section. 

The critical field $H_{c} $ is determined by the maximum of
$\tilde{H}(q)$ as the function of $q$. The FFLO state arises when the
maximum of $H$ occurs at finite wave vector
\begin{eqnarray}
&&q_{0} ^{2} =\\ 
&&\frac{K_{03} \langle|\psi|^{2}  v_{z} ^{2} \rangle
-\frac{3K_{05} \langle|\psi|^{2} v_{\perp} ^{2} v_{z} ^{2} \rangle}{4\lambda^{2} }
-\frac{K_{15} \Re(\langle\psi^* v_{\perp} ^{2} \rangle\langle\psi
v_{z} ^{2} \rangle)}{8\tau\lambda^{2} }}
{\frac{K_{05} \langle|\psi|^{2} v_{z} ^{4} \rangle}{2}+
\frac{K_{15} |\langle \psi v_{z} ^{2} \rangle|^{2} }{4\tau}}.\nonumber
\label{e26}
\end{eqnarray}
The FFLO state appears with the sign change of the coefficient at 
$q^{2}$ in Eq.~ (\ref{e25}) and it exists at
\begin{equation}
K_{03} ~<~\frac{3K_{05} \langle|\psi|^{2} v_{\perp} ^{2} v_{z} ^{2} \rangle
+K_{15} \Re(\langle\psi^* v_{\perp} ^{2} \rangle\langle\psi
v_{z} ^{2} \rangle)/2\tau}{4\lambda^{2} \langle|\psi|^{2}  
v_{z} ^{2} \rangle}.
\label{e27}
\end{equation}

The type of transition changes from the second to the first order with
the sign change of the coefficient $\tilde B$ at $|\varphi_0|^{4} $ 
in Eq.~(\ref{e21}).  So, the first order transition from normal to Abrikosov 
vortex lattice state persists when $\tilde B_0<0$, that is in the region of 
validity of the following inequality
\begin{equation} 
K_{03}
<
\frac{K_{04}}{2\tau \langle|\psi|^4\rangle}
+
\frac{K_{05}}{2\lambda^2}
\frac{\langle|\psi|^4 v_\perp^2\rangle}{\langle|\psi|^4\rangle}
-\frac{5 K_{06}}{16\tau\lambda^2}
\frac{\langle|\psi|^2 v_\perp^2\rangle}
{\langle|\psi|^{4} \rangle}.
\label{e28}
\end{equation}

To see explicitly the role of orbital effects and the impurities in 
the formation of FFLO state and the change of the transition type, 
let us
look on them separetely.

In pure limit $\tau\to\infty$,
the two inequalities (\ref{e27}) and (\ref{e28}) take the much 
simpler form, where FFLO exists at
\begin{equation}
K_{03} ~<~\frac{3K_{05} \langle|\psi|^{2} 
v_{\perp} ^{2} v_{z} ^{2} \rangle
}{4\lambda^{2} \langle|\psi|^{2}  v_{z} ^{2} \rangle}
\label{e29}
\end{equation}
and the first order transition to vortex lattice state at
\begin{equation}
K_{03} ~<~\frac{K_{05} \langle|\psi|^{4} v_{\perp} ^{2} \rangle}
{2\lambda^{2} \langle|\psi|^{4} \rangle}.
\label{e30}
\end{equation}
In quasi two-dimensional case, we deal in first approximation
with cylindrical Fermi surface: $\psi_k=\sqrt{2}(\hat k_x^2-\hat 
k_y^2)$ 
and $v_\perp\simeq v$ is constant. Then, the inequalities look even 
simpler:
FFLO state exists at
\begin{equation}
K_{03} ~<~\frac{3K_{05} v^{2}  }{4\lambda^{2} }
\label{e31}
\end{equation}
and the first order transition to vortex lattice state at
\begin{equation}
K_{03} ~<~\frac{K_{05} 
v^{2} } {2\lambda^{2} }.
\label{e32}
\end{equation}
The comparison of these two expressions makes evident that the 
critical point where the second order transition from normal metal to 
superconductor transforms to the first order one lies at higher temperature
than the point at which FFLO state arises. The resulting phase 
diagram is qualitatively shown on Fig.~\ref{fig3}(c). It corresponds to the 
experimental observation in $CeCoIn_{5} $ \cite{Bian03}.

This conclusion is just the opposite to the case of s-wave superconductivity 
in isotropic metal considered in previous subsection. It is obvious that 
this difference has pure geometrical origin (order parameter and the Fermi 
surface anisotropy) and does not originate from the difference in applied 
theoretical approaches: gradient expansion in \cite{Houz} and nonpertubative 
treatment \cite{Ik}.

In absence of orbital effect ($\lambda\to\infty$), the FFLO state exists at 
$K_{03}<0$. In the pure limit ($\tau\to\infty$), 
it gives therefore higher critical field than the critical field 
corresponding to the transition from normal to uniform 
superconducting state which also changes its type at the same place.
In this pure, paramagnetic limit, $K_{03}=K_{30}$ and $\tilde H_0(T)=H_0(T)$.
Therefore, the tricriical point defined by $K_{30}=0$ along the critical line 
$H_0(T)$ is still $(T^*,H^*)$ for d-parining. The actual structure of the FFLO 
state which is realized just below the critical line, at $T<T^*$, was considered 
in Refs.~\cite{MakiWon,Sam,Vor, Agt} for quasi two-dimensional d-wave superconductors. 
It was shown that, close to the tricritical point, the second order transition 
takes place from normal state to sinusoidally modulated state, with the direction of 
modulation parallel to the conducting planes and along the nodes of the order parameter. 
The resulting phase diagram is illustrated in Fig.~\ref{fig3}(a). In our 
analysis, we considered the modulation along the applied field. It 
yields $\tilde B_{\text{exp}}(q_0)=-3K_{03}/8$ and $\tilde 
B_{\text{sin}}(q_0)=-K_{03}/16$, that is $\tilde 
B_{\text{exp}}(q_0)>\tilde B_{\text{sin}}(q_0)>0$ in the region of 
existence of FFLO state. Therefore, we obtain the same topology for 
the phase diagram as shown in Fig.~\ref{fig3}(a).

In absence of orbital effect ($\lambda\to\infty$), but with some disorder, one can 
check that, at the point where FFLO instability takes place, defined by $K_{03}=0$, 
$\tilde B_0$ is still positive. Therefore, the phase transition from 
normal to uniform superconducting state is still of the second order. 
Thus the influence of impurities is also opposite to the case of s-pairing in
isotropic metal. If we consider the modulation along the applied 
field, we find $\tilde B_{\text{exp}}(q_0)=-3K_{03}/8-K_{04}/8\tau$ 
and $\tilde B_{\text{sin}}(q_0)=-K_{03}/16-3K_{04}/16\tau$. Thus, 
both are positive, due to negative value of $K_{04}$ at the 
tricritical point and transition from normal to FFLO state remains of 
the second type, while the sinusoidal (exponential) modulation is 
favored at $K_{03}<K_{04}/5\tau$ ($K_{04}/5\tau<K_{03}<0$). The 
corresponding phase diagram is qualitatively shown in 
Fig.~\ref{fig3}(b). It has the same topology as the phase diagram 
given for modulation along the nodes of the order parameter in the 
presence of impurities in Ref.~\cite{Agt}.

\begin{figure}[tbp]
\includegraphics[scale=1]{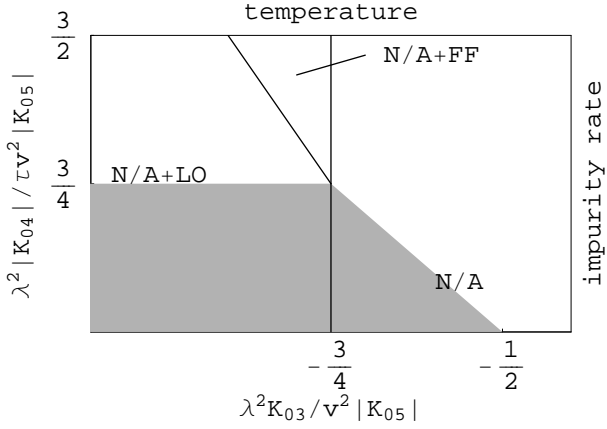}
\caption{
This figure shows the nature of the superconducting state which is 
realized just below the upper critical line, as well as the type of the 
transition, for quasi two-dimensional d-wave superconductors, at large 
paramagnetic, small orbital and impurity effects, and for temperatures 
close to the tricritical temperature $T^*$. (See also the legend of 
Fig.~\ref{fig2}).
}
\label{fig4}
\end{figure}

When both orbital effect and impurities are present and small, and 
assuming an almost cylindrical Fermi surface, we can obtain the more 
precise following picture for the phase diagram. In the leading order 
in $1/\lambda^2,1/\tau$, one finds that the equations defined by 
(\ref{e27}) and (\ref{e28}) cross at 
$K_{04}/2\tau=3v^{2} K_{05}/8\lambda^2$. When
\begin{equation}
\frac{1}{\tau}
<
\frac{3v^{2} }{4\lambda^2}
\frac{K_{05}}{K_{04}},
\label{eq44}
\end{equation}
transition from normal to usual superconducting vortex state turns 
from second to first order when temperature is lowered. The modulated 
FFLO state may exist as a stable or metastable state below the first order 
transition line, with eventually even larger critical field at lower 
temperatures (Fig.~\ref{fig3}(c)). On the other hand, when inequality 
(\ref{eq44}) is reversed, the impurity effect dominates: as the temperature 
is lowered, the transition from normal to superconducting state is changed into 
the transition from normal to exponentially modulated state at 
$K_{03}<3v^2K_{05}/4 \lambda^2$. Then, it is changed into the 
transition into sinusoidally modulated state at $\tilde 
B_{\text{exp}}(q_0)>\tilde B_{\text{sin}}(q_0)$. With 
Eqs.~(\ref{eBexpd}), (\ref{eBsind}), (\ref{e28}), and appropriate 
averaging over Fermi surface, one finds that this occurs at:
\begin{equation}
K_{03}<
\frac{K_{04}}{5 \tau}
+
\frac{3v^2K_{05}}{5 \lambda^2}.
\end{equation}
In this region, $\tilde B_{\text{exp}}(q_0)$ and $\tilde 
B_{\text{sin}}(q_0)$ remain positive, therefore the critical line 
remains of the second order. These results are summarized in the 
Fig.~\ref{fig4}.

From the previous discussion, one can guess that our Ansatz 
(\ref{e21}) is not the most general. Indeed, in the presence of orbital 
effect, one should also consider order parameters in the form
$$
\Delta=f(z)(A_n\varphi_n+A_{n+4}\varphi_{n+4}+\ldots),
$$ 
where $n>0$. Therefore, the FFLO phases illustrated in 
Figs.~\ref{fig3} and \ref{fig4} may compete with phases corresponding
to order parameter with higher Landau level $n>0$.  This problem is
reserved for further study.

Note that these results are very sensitive to the shape of the Fermi 
surface. In particular, different topology
for the phase diagram would be obtained for anisotropic d-wave 
superconductor with elliptic Fermi surface.

\section{FFLO instability in disordered s-wave superconductor}

According to Ref.~\cite{Asl}, instability toward FFLO state formation 
is always present in s-wave superconductors in the paramagnetic 
limit. In particular, assuming that the transition from normal to 
FFLO state is of the second type, such instability occurs at 
$T^*=0.56 T_c$ in clean systems and at vanishingly small temperature 
$T_d^*\simeq -\Delta_0/(2\ln \tau \Delta_0)$ in dirty ones. 

On the other hand, a large orbital effect is detrimental to FFLO 
instability, as shown in clean s-wave superconductors in 
Ref.~\cite{Gru}. Indeed, it was shown there that the FFLO instability 
only takes place when the paramagnetic effect, characterized by a 
Maki parameter $\alpha_M^c>1.8$, is strong enough.

In this section, we adress the question of FFLO instability in s-wave 
superconductors with disorder. In particular, we find that the second 
order transition toward FFLO state exists for any disorder provided 
that the orbital effect remains small enough. In the dirty limit 
($\tau \Delta_0 \ll 1$), the Maki parameter characterizing its 
strength must be very large: $\alpha_M^d> -1/(\tau\Delta_0\ln \tau 
\Delta_0)$.

As we discussed in the previous sections, transition from normal to 
conventional superconducting state may change its type. In 
particular, in dirty systems, such change of the transition type was 
shown to take place below some critical temperature as soon as 
$\alpha_M^d> 1$.\cite{maki66} Therefore, the FFLO state may either 
exist as a metastable state below the first order transition line 
into conventional superconducting state or take place by first order 
transition with even higher critical field. We do not discuss the 
question of the type of the transition in the present section.

Let us now derive the result.     

At the second order transition in superconducting state, the 
linearized self-consistency equation (\ref{B-gap}) is
\begin{equation} \label{e4-1}
\Delta \ln\frac{T}{T_c}
+
2 \pi T \Re
\left[
\sum_{\nu \geq 0}
\left(
\frac{1}{\omega_\nu}
-
\frac{ \langle {\cal L}_\nu \rangle }{1-\frac{1}{2\tau}\langle {\cal 
L}_\nu \rangle} 
\right)
\right]
\Delta
=0,
\end{equation}
where the differential operator $\langle {\cal L}_\nu \rangle$ is:
\begin{equation}\label{e4-2}
\langle {\cal L}_\nu \rangle
=
\int
\frac{d \Omega_{\bf{v}}}{4\pi}
\frac{1}{\tilde{\Omega}_\nu+i\frac{\bf{vD}}{2}}.
\end{equation}

The  most general form of the solution for the gap at the second 
order transition is $\Delta=\varphi_{0}(x,y)e^{iqz}$, 
where $q$ is the FFLO modulation vector and $\varphi_{0}$ is the 
Abrikosov vortex lattice formed of lowest level Landau functions. 
Using the identity $1/X=\int_0^\infty ds e^{-sX}$ and the properties 
of Landau functions, we find that the operator $\langle {\cal L}_\nu 
\rangle$
applied to $\Delta$ yields the eigenvalue:
\begin{eqnarray}\label{e4-3}
\langle {\cal L}_\nu (q) \rangle 
=
\int_0^\infty ds
\exp (-s \tilde{\Omega}_\nu) 
\int_0^1 du
\qquad \qquad 
\\
\quad
\times
\exp \left(-\frac{\pi H v_F^2}{8 \Phi_0}s^2(1-u^2)\right)
\cos\left(\frac{suv_Fq}{2}\right)
.
\nonumber 
\end{eqnarray}

At $q=0$, Eqs.~(\ref{e4-1}),(\ref{e4-3}) yield the second order 
critical line $H_{c2}(T)$ for transition into usual superconducting 
vortex lattice state. In particular, in the paramagnetic limit the 
upper critical field is $H_p^{(2)}=\Delta_0/(2\mu)=H_p/\sqrt{2}$ and 
it does not depend on the disorder. In the clean, orbital limit, the 
upper critical field is $H_{c20}^{c}=(\gamma e^2/2\pi) \Phi_0 
\Delta_0^2 /v_F^2$. At finite temperature and/or intermediate disorder, 
$H_{c2}$ must be found numerically. 

In the dirty limit, the equations determining the upper critical 
field simplify greatly. Indeed, integration on $s$ in Eq.~(\ref{e4-3}) is 
cut off by impurity scattering time, $s \lesssim \tau$. Thus, 
assuming that $H \ll \Phi_0/(v_F \tau)^2$ (as can be checked 
consistently later), we may expand the second exponential in Eq.~(\ref{e4-3}) 
and perform the integration explicitly:
\begin{equation}\label{e4-4}
\langle {\cal L}_\nu (0) \rangle
\simeq 
\left(
\frac{1}{\tilde{\Omega}_\nu}
-\frac{\pi H v_F^2}{6 \Phi_0 \tilde{\Omega}_\nu^3} 
\right)
.
\end{equation}
Thus, we obtain the implicit equation for $H_{c2}$ in the dirty limit
\cite{K.Maki}:
\begin{equation}\label{e4-5}
\ln\frac{T}{T_{c}}
=\Re\left[
\Psi\left(\frac{1}{2}\right)
-\Psi\left(\frac{1}{2}+i\frac{\mu H_{c2}}{2\pi 
T}+\frac{DH_{c2}}{2\Phi_0T}\right)
\right],
\end{equation}
where $D=v_F^2\tau/3$ is the diffusion constant. In particular, the 
critical field at zero temperature,
\begin{equation}\label{e4-6}
H_{c2}(0)=
\frac{H_{c20}^{d}}{\sqrt{1+\alpha_M^{d2}}},
\end{equation}
interpolates between $H_p^{(2)}$ in the paramagnetic limit and the 
upper critical field in the orbital, dirty limit, 
$H_{c20}^{d}=\Phi_0T_c/2\gamma D$, with the Maki parameter in the 
dirty limit defined as: $\alpha_M^d=\sqrt{2}H_{c20}^{d}/H_p=\mu 
\Phi_0/\pi D$.

In general, the critical field defined by Eq.~(\ref{e4-1}) also 
depends on $q$: $H=H(q)$. The actual critical field corresponds to 
the maximal value of $H(q)$ with respect to $q$. When it is obtained for 
$q\neq 0$, second order transition into FFLO state is realized. Along the 
critical line $H(q=0)$ at given impurity rate and Maki parameter, the triple 
point below which such transition may occur is defined by
$\partial H/\partial (q^2)|_{q=0}=0$. (One could check that $\partial 
H/\partial q|_{q=0}=0$ is always true.)
In order to obtain $\partial H/\partial (q^2)|_{q=0}=0$, one can 
expand Eq.~(\ref{e4-1}) up to the second order in $q$, and obtain:
\begin{eqnarray}
H&=&H_{c2}(T)
+ \Lambda  q^2+ {\cal O} (q^4)
\\
\nonumber
\Lambda&=&
\frac{(2 \pi T)^2}{\mu \Re\left[\zeta 
\Psi_1\left(\frac{1}{2}+\frac{\mu H \zeta}{2\pi T}\right) \right]}
\Re \sum_{\nu \geq 0}
\left.
\frac{ \partial \langle {\cal L}_\nu (q) \rangle /\partial (q^2)  }
{(1-\frac{1}{2\tau} \langle {\cal L}_\nu (q) \rangle )^2}
\right|_{q=0}.
\end{eqnarray}
where $\zeta=1/\alpha_M^d+i$. Making use of Eq.~(\ref{e4-3})
and integration by  part, one finally obtains the condition  
$\partial H/\partial (q^2)|_{q=0}\propto \Lambda=0$ in the form:
\begin{equation} \label{e4-7}
0
=
\Re \sum_{\nu \geq 0}
\frac{ \tilde{\Omega}_\nu^{-1}- \langle {\cal L}_\nu (0) \rangle  }
{(1-\frac{1}{2\tau} \langle {\cal L}_\nu (0) \rangle )^2}.
\end{equation}
In particular, at zero temperature this equation defines the minimal 
Maki parameter which allows existence of FFL0 state. In the dirty limit, 
Eq.~(\ref{e4-7}) is easily integrated at zero temperature:
\begin{eqnarray} \nonumber
0&=&
\Re
\int_0^\infty d\omega
\frac{1}{\omega+\frac{1}{2\tau}+i \mu H}
\frac{1}{(\omega+\frac{\pi D H}{\Phi_0}+i \mu H)^2}
\\
&\simeq&
\frac{\sqrt{1+(\alpha_M^d)^2}}{\alpha_M^d (\tau \Delta_0)^2}
\left(
\frac{1}{\alpha_M^d}
-
\tau \Delta_0
\ln
\frac{1}{\tau\Delta_0}
\right)
\end{eqnarray}
It yields the critical Maki parameter above which FFLO 
instability exists:
\begin{equation}
\alpha_M^d
\simeq
\frac{1}{ 
	\tau\Delta_0
	\ln\frac{1}{\tau\Delta_0}
	 }.
\end{equation}
Therefore, FFLO instability is present in dirty superconductors 
provided that the orbital effect is small enough.

\section{Conclusion}

In conclusion, we derived microscopically the generalized 
Ginzburg-Landau free energy functional which is adequate to describe 
conventional and unconventional singlet superconductors in the
presence of paramagnetic, orbital and impurity effects. This free 
energy was used to predict the superconducting phase diagrams of 
three-dimensional s-wave superconductors and quasi-two-dimensional 
d-wave superconductors under magnetic field perpendicular to the 
conducting layers. These phase diagrams prove to be quite different and
to be very sensitive to geometrical effects such as the nature of the 
order parameter and the shape of the Fermi surface. 
In particular, we found that impurities tend to favor the transition 
from normal state to the Abrikosov vortex lattice state, with the change of the 
transition type as the temperature is lowered in the s-wave case, 
while they tend to favor the transition from normal state to the FFLO 
state with vortex lattice plus additional modulation of the order 
parameter along the field direction in the d-wave case. We also found 
that the orbital effect acts in the opposite direction. That is, it 
tends to favor transition from normal to the FFLO state in s-wave 
case, while it tends to favor the transition from normal state to the 
vortex lattice state with the change of the transition type in the 
d-wave case. 

In addition, we determined the criterion for instability toward non 
uniform superconducting state in s-wave superconductors in the dirty 
limit.

\appendix

\section{Free energy functional in superconductor doped by impurities}

The free energy functional expanded over the order parameter  
for superconducting state with pairing interaction
\begin{equation}
V({\bf k},{\bf k}')=-V\psi(\hat{\bf k})\psi(\hat{\bf k}')
\label{eA1}
\end{equation}
has the following form
\begin{eqnarray} 
&&F=\sum_{{\bf q}}\frac{|\Delta_{{\bf q}} |^{2} }{V}-\nonumber\\
&&T\sum_{\omega} \sum_{{\bf k}{\bf k}'{\bf q}'}  G^{\omega}  ({\bf 
k},{\bf
k}')
G^{-\omega}(-{\bf k}+{\bf q}, -{\bf k}'+{\bf q}')
\Delta^{*} _{{\bf k},{\bf q}}\Delta_{{\bf k}',{\bf q}'}  \nonumber\\
&&+\frac{T}{2}\sum_{\omega} \sum_{{\bf
q}_{1} - {\bf q}_{2} ={\bf q}_{4} -{\bf q}_{3} } 
\sum_{{\bf k}{\bf p}{\bf l}{\bf m}}  G^{\omega}  ({\bf k},{\bf p})
\Delta^{*} _{{\bf k},{\bf q}_{1} } \nonumber\\
&&G^{-\omega}  (-{\bf k}+{\bf q}_{1}  ,-{\bf l}+{\bf
q}_{1} )\Delta_{{\bf l},{\bf q}_{2} } \nonumber\\
&&G^{\omega}  ({\bf m}-{\bf q}_{1} +{\bf q}_{2} ,{\bf l}-{\bf
q}_{1} +{\bf q}_{2} ) \Delta^{*} _{{\bf m},{\bf q}_{3} } \nonumber\\
&&G^{-\omega}  (-{\bf m}+{\bf q}_{4} , -{\bf p}+{\bf q}_{4} )
\Delta_{{\bf p}, {\bf q}_{4}} .
\label{eA2}
\end{eqnarray}
Here, the order parameter is given by Fourier transformation of Eqn
(\ref{e1})
$$
\Delta_{{\bf k},{\bf q}}=\psi(\hat{\bf k})\Delta_{{\bf q}} =\int
d^{3} r\exp(-i{\bf q}{\bf r})\Delta({\bf k},{\bf r}),
$$
and $G^{\omega}  ({\bf k},{\bf p})$ is exact electron Green
function in normal metal with arbitrary configuration of impurities. 
Averaging of the free energy over impurity configurations demands
calculation of averages of vertices \cite{Gor60}
\begin{eqnarray}
&&A_{{\bf k},{\bf q}} =\nonumber\\
&&\sum_{{\bf k}'{\bf q}'} 
\overline {G^{\omega}  ({\bf k},{\bf k}')G^{-\omega}(-{\bf k}+{\bf q},
-{\bf k}'+{\bf q}')\Delta_{{\bf k}',{\bf q}'} }
\label{eA3}
\end{eqnarray}
obeying of equation
\begin{eqnarray}
&&A_{{\bf k},{\bf q}} =\sum_{{\bf k}'{\bf q}'} 
\overline {G^{\omega}  ({\bf k},{\bf k}')}~\overline {
G^{-\omega}(-{\bf k}+{\bf q}, -{\bf k}'+{\bf q}'}) \nonumber\\
&&[\Delta_{{\bf k}',{\bf q}'} +nu^{2} \sum_{{\bf p}} 
A_{{\bf p},{\bf q}'} ],
\label{eA4}
\end{eqnarray}
where $nu^{2} =1/2\pi N_{0} \tau$, $n$ is impurity concentration, $u$
is the amplitude of scattering, and $\tau$ is mean free time of
scattering of quasiparticles.  Substituting the Green functions by 
its average
\begin{equation}
\overline {G^{\omega}  ({\bf k},{\bf p})} = G^{\omega}  ({\bf
k})\delta({\bf k}-{\bf p}),
\label{eA5}
\end{equation}
\begin{equation}
G^{\pm\omega}  ({\bf k})= \frac{1}{\pm i(\tilde\omega_{\nu} -i\mu
H)-\xi({\bf k})}
\label{eA6}
\end{equation}
we obtain from (\ref{eA4})
\begin{equation}
A_{{\bf k},{\bf q}} =G^{\omega}  ({\bf k}) G^{-\omega}(-{\bf
k}+{\bf q})\bar\Delta_{{\bf k},{\bf q}} ,
\label{eA7}
\end{equation}
where
\begin{equation}
\bar\Delta_{{\bf k},{\bf q}} =
\Delta_{{\bf k},{\bf q}} +nu^{2} \sum_{{\bf p}}  A_{{\bf p},{\bf
q}} ,
\label{eA8}
\end{equation}
and
\begin{equation}
\sum_{{\bf k}} A_
{{\bf k},{\bf q}} =\frac{\sum\limits_{{\bf k}} 
G^{\omega}  ({\bf k}) G^{-\omega}(-{\bf k}+{\bf q}) \Delta_{{\bf
k},{\bf q}} } {1-nu^{2} \sum\limits_{{\bf k}} G^{\omega}  ({\bf k})
G^{-\omega}(-{\bf k}+{\bf q})}.
\label{eA9}
\end{equation}
Then, following the procedure developed in \cite{Gor60}, after the 
averaging
of free energy (\ref{eA2}) we obtain
\begin{equation}
F=F_{2} +F_{4} 
\label{eA10}
\end{equation}
where
\begin{eqnarray}
&&F_{2} =\sum_{{\bf q}}\frac{|\Delta_{{\bf q}} |^{2} }{V}-\nonumber\\
&&T\sum_{\omega} \sum_{{\bf k}{\bf q}} G^{\omega}  ({\bf k})
G^{-\omega}(-{\bf k}+{\bf q}) \Delta^{*} _{{\bf k},{\bf
q}}\bar\Delta_{{\bf k},{\bf q}} 
\label{eA11}
\end{eqnarray}
\begin{eqnarray}
&&F_{4} =\frac{T}{2}\sum_{\omega}  \sum_{{\bf
q}_{1} - {\bf q}_{2} ={\bf q}_{4} -{\bf q}_{3} } 
\left\{\sum_{{\bf k}}  
G^{\omega}  ({\bf k}) \bar\Delta^{*} _{{\bf k},{\bf
q}_{1} } \right.\nonumber\\
&&\left.G^{-\omega}  (-{\bf k}+{\bf q}_{1} )\bar\Delta_{{\bf
k},{\bf q}_{2} }  G^{\omega}  ({\bf k}-{\bf q}_{1} +{\bf q}_{2} )
\bar\Delta^{*} _{{\bf k},{\bf q}_{3} } \right.\nonumber\\
&&\left.G^{-\omega}  (-{\bf k}+{\bf q}_{4} )\bar\Delta_{{\bf k}, {\bf
q}_{4}} \right.\nonumber\\.  &&\left.+nu^{2} \sum_{{\bf
k}}G^{\omega}  ({\bf k}) \bar\Delta^{*} _{{\bf k},{\bf
q}_{1} } \right.\nonumber\\
&&\left.G^{-\omega}  (-{\bf k}+{\bf q}_{1} )\bar\Delta_{{\bf k},
{\bf q}_{4}} G^{-\omega}  (-{\bf k}+{\bf q}_{4} )\right.\nonumber\\
&&\left.\sum_{{\bf k}}G^{-\omega}  (-{\bf k}+{\bf q}_{1} )
\bar\Delta_{{\bf k},{\bf q}_{2} }G^{-\omega}  (-{\bf k}+{\bf
q}_{4} )\right.\nonumber\\
&&\left.\bar\Delta^{*} _{{\bf k},{\bf q}_{3} }  G^{\omega}  ({\bf
k}-{\bf q}_{1} +{\bf q}_{2} )\right.\nonumber\\
&&\left.+nu^{2} \sum_{{\bf k}}G^{-\omega}  (-{\bf k}+{\bf q}_{4} )
\bar\Delta_{{\bf k}, {\bf q}_{4}} \right.\nonumber\\
&&\left.G^{\omega}  ({\bf k})\bar\Delta^{*} _{{\bf k},{\bf q}_{3} } 
G^{\omega}  ({\bf k}-{\bf q}_{1} +{\bf q}_{2} )\right.\nonumber\\
&&\left.\sum_{{\bf k}}G^{-\omega}  (-{\bf k}+{\bf q}_{1} )
\bar\Delta^{*} _{{\bf k},{\bf q}_{1} } \right.\nonumber\\
&&\left.G^{\omega}  ({\bf k})\bar\Delta_{{\bf k},{\bf
q}_{2} } G^{\omega}  ({\bf k}-{\bf q}_{1} +{\bf q}_{2} )\right\}
\label{eA12}
\end{eqnarray}

The further calculations are different for different superconducting 
states. 
Expanding the quadratic in respect of the order parameter terms up to 
forth
order in respect of ${\bf q}$ and performing $\xi$ integration we 
obtain for
the case of s-pairing
\begin{eqnarray}
&&\sum_{{\bf k}} G^{\omega}  ({\bf k}) G^{-\omega}(-{\bf k}+{\bf q})
\Delta^{*} _{{\bf k},{\bf q}}\bar\Delta_{{\bf k},{\bf q}}  \nonumber\\
&&=\frac{\pi N_{0} }{\Omega_{\nu} }\left\{ 1-\frac{\langle({\bf q}{\bf
v})^{2} \rangle}{4\Omega_{\nu} \tilde\Omega_{\nu} }+ 
\frac{\langle({\bf q}{\bf
v})^{4} \rangle} 
{16\Omega_{\nu} \tilde\Omega_{\nu} ^{3} }\right.\nonumber\\
&&\left.+\frac{(\langle({\bf q}{\bf
v})^{2} \rangle)^{2} }{32\tau\Omega_{\nu} ^{2} \tilde\Omega_{\nu} ^{3} }
\right\}|\Delta_{{\bf q}}|^{2} .
\label{eA13}
\end{eqnarray}
Corresponding expression for superconducting order parameter
transforming according to non-identity representation even in respect
of ${\bf k}$ is
\begin{eqnarray}
&&\sum_{{\bf k}} G^{\omega}  ({\bf k}) G^{-\omega}(-{\bf k}+{\bf q})
\Delta^{*} _{{\bf k},{\bf q}}\bar\Delta_{{\bf k},{\bf q}}  \nonumber\\
&&=\frac{\pi N_{0} }{\tilde\Omega_{\nu} }\left\{ 1- \frac{\langle({\bf
q}{\bf v})^{2} |\psi(\hat{\bf k})|^{2} \rangle} 
{4\tilde\Omega_{\nu} ^{2} }+
\frac{\langle({\bf q}{\bf v})^{4}  |\psi(\hat{\bf k})|^{2} \rangle}
{16\tilde\Omega_{\nu} ^{4} }\right.\nonumber\\
&&\left.+\frac{|\langle({\bf q}{\bf
v})^{2} \psi(\hat{\bf
k})\rangle|^{2} }{32\tau\Omega_{\nu} \tilde\Omega_{\nu} ^{4} }
\right\}|\Delta_{{\bf q}}|^{2} .
\label{eA14}
\end{eqnarray}
Here
\begin{equation}
\Omega_{\nu} =\omega_{\nu} -i\mu H,~~ 
\tilde\Omega_{\nu} =\tilde\omega_{\nu} -i\mu H,~~
\tilde\omega_{\nu} =\omega_{\nu} +\frac{sign ~\omega_{\nu} }{2\tau}.
\label{eA15}
\end{equation}

Substituting (\ref{eA13}) and (\ref{eA14}) in (\ref{eA11}) and 
performing
Fourier transformation to the coordinate space 
(accompanied by substitution ${\bf q}\to{\bf D}=-i\nabla+(2\pi
/\Phi_{0} ){\bf A}$) we come to the quadratic terms in the Eqns
(\ref{e2}) and (\ref{e7}) correspondingly.

In terms of the forth order in respect of
the order parameter one should calculate $\bar\Delta_{{\bf k},{\bf
q}} $ up to the second order in ${\bf q}$.  For the s-pairing state it
is
\begin{equation}
\bar\Delta_{{\bf k},{\bf 
q}} =\frac{\tilde\Omega_{\nu} }{\Omega_{\nu} }
\left(1-\frac{\langle({\bf q}{\bf
v})^{2} \rangle}{8\tau\Omega_{\nu} \tilde\Omega_{\nu} ^{2} }\right)
\Delta_{{\bf q}} 
\label{eA16}
\end{equation}
and for a non-identity representation
\begin{equation}
\bar\Delta_{{\bf k},{\bf q}} = \left(\psi(\hat{\bf k})
-\frac{\langle({\bf q}{\bf
v})^{2} \psi(\hat{\bf 
k})\rangle}{8\tau\Omega_{\nu} \tilde\Omega_{\nu} ^{2} }\right) 
\Delta_{{\bf
q}} .
\label{eA17}
\end{equation}
It is easy to check that in the latter case in terms of the forth 
order
in respect of the order parameter and up to the second order in ${\bf
q}$ one can put just $\bar\Delta_{{\bf k},{\bf q}} = \psi(\hat{\bf k})
\Delta_{{\bf q}} $.

Substituting (\ref{eA16}) and (\ref{eA17}) in (\ref{eA12}), expanding
the Green functions up to the second order in ${\bf q}$, performing
the integration over ${\xi}$ and Fourier transformation to the
coordinate space 
we come to the quartic terms in the Eqns (\ref{e2}) and (\ref{e7})
correspondingly.

\section{Free energy functional derived from Eilenberger equations}

In this appendix we propose another method to derive the 
superconducting free energies (\ref{e2}) and (\ref{e7}) which are 
introduced in Sec.~\ref{sec2}.

The quasiclassical theory of superconductivity forms a convenient 
framework to study conventional and unconventional superconductors 
in the presence of magnetic fields or impurities.\cite{kopnin} In 
this theory, the superconducting gap is related to the anomalous 
function $f_\nu(\hat{\bf k},{\bf r})$ through:
\begin{equation} \label{B-gap}
\Delta({\bf r})
=
\pi T V \sum_\nu
\langle 
\psi(\hat{\bf k})^*
f_\nu(\hat{\bf k},{\bf r})
\rangle_{\hat{\bf k}},
\end{equation}
where the brackets stand for the averaging over the Fermi surface 
labelled by $\hat{\bf k}$, $V$ and $\psi(\hat{\bf k})$ define the 
pairing interaction (\ref{eA1}).
The anomalous function is determined by the set of Eilenberger 
equations:
\begin{equation} \label{B-eilenberger}
\frac{i}{2} {\bf vD}
f_\nu
+
(	
	\Omega_\nu 
	+\frac{1}{2\tau} \langle g_\nu \rangle
) 
f_\nu
=
\left(
\psi \Delta
+
\frac{1}{2\tau} \langle f_\nu \rangle
\right) 
g_\nu,
\end{equation}
where 
\begin{equation} \label{eB3}
g_\nu=\text{sign}(\omega_\nu) \sqrt{1-f_\nu f^\dagger_\nu}
\quad \text{and} \quad
f^\dagger_\nu=-f^*_{-(\nu+1)}.
\end{equation}
The magnetic field ${\bf H}=\text{rot} {\bf A}$ is combined with 
gradient in ${\bf D}=-i\nabla+(2\pi/\Phi_{0} ){\bf A}$, and 
$\Omega_\nu=\omega_\nu-i \mu H$ where $\omega_\nu$ is a Matsubara 
frequency.

Near the second order transition from normal to superconducting 
state, the order parameter $\Delta$ is vanishingly small. Moreover, 
we assume that the order parameter is slowly varying on the scale of 
the superconducting coherence length. Then we may expand the 
selfconsistency equation (\ref{B-gap}) up to the third order terms in 
the gap, and the fourth order terms in the gradient expansion. 

In the following we proceed separately for the cases of identity and 
non-identity representation.

\subsection{Identity representation}

For simplicity, we only consider identity represenation with  
$\psi=1$.

The set of Eilenberger equations can be expanded perturbatively in 
the gap. In this expansion, $g_\nu$ only contains even terms: 
$g_\nu=g_\nu^{(0)}+g_\nu^{(2)}+\dots$, while $f_\nu$ only contains 
odd terms: $f_\nu=f_\nu^{(1)}+f_\nu^{(3)}+\dots$.

In the zeroth order in $\Delta$, at $\omega_\nu>0$, we find 
$g_\nu^{(0)}=1$.

In the first order in $\Delta$, $f_\nu^{(1)}$ is the solution of the 
linearized differential equation (\ref{B-eilenberger}):
\begin{equation}
\frac{i}{2} {\bf vD}
f_\nu^{(1)}
+
\tilde{\Omega}_\nu 
f_\nu^{(1)}
=
\Delta
+
\frac{1}{2\tau} \langle f_\nu^{(1)} \rangle,
\end{equation}
where $\tilde{\Omega}_\nu=\Omega_\nu+1/2\tau$. Expressing $\langle 
f_\nu^{(1)} \rangle$ in terms of $\Delta$, we  get:
\begin{equation}
\langle 
f_\nu^{(1)}
\rangle
=
\frac
	{\langle {\cal L}_\nu \rangle}
	{1-\frac{1}{2\tau} \langle {\cal L}_\nu \rangle}
\Delta,
\end{equation}
where ${\cal L}_\nu=(\tilde{\Omega}_\nu+i{\bf vD}/2)^{-1}$.
Expanding up to the fourth order terms in the gradient expansion, we 
find:
\begin{equation} \label{B-f1-Id}
\langle 
f_\nu^{(1)}
\rangle
=
\frac{\Delta}{\tilde{\omega}_\nu} 
-\frac{\langle ({\bf vD})^2\rangle 
\Delta}{4\tilde{\omega}_\nu^2\tilde{\Omega}_\nu}
+
\frac{\langle ({\bf vD})^4\rangle 
\Delta}{16\tilde{\omega}_\nu^2\tilde{\Omega}_\nu^3}
+\frac{\langle ({\bf vD})^2\rangle^2 
\Delta}{32\tau\tilde{\omega}_\nu^3\tilde{\Omega}_\nu^3}
,
\end{equation}
where $\tilde{\omega}_\nu=\omega_\nu+1/2\tau$.

In the second order in $\Delta$, we find from Eq.~(\ref{eB3}): 
$g_\nu^{(2)}=-f_\nu^{(1)}f_\nu^{\dagger(1)}/2$.

In the third order in $\Delta$, on gets from 
Eq.~(\ref{B-eilenberger}) that $f_\nu^{(3)}$ is the solution of the 
linear differential equation:
\begin{eqnarray}
\frac{i}{2} {\bf vD}
f_\nu^{(3)}
+
\tilde{\Omega}_\nu 
f_\nu^{(3)}
=
\Delta g_\nu^{(2)}
+
\frac{1}{2\tau} \langle f_\nu^{(3)} \rangle
\nonumber \\
+
\frac{1}{2\tau} 
\left(
	\langle f_\nu^{(1)} \rangle g_\nu^{(2)}
	-
	\langle g_\nu^{(2)} \rangle f_\nu^{(1)}
\right).
\end{eqnarray}
We can now express $\langle f_\nu^{(3)} \rangle$ in terms of $\Delta$ 
and we make an expansion up to terms of the second order in the 
gradient expansion. We obtain:
\begin{eqnarray} \label{B-f3-Id}
\langle f_\nu^{(3)} \rangle
=
-\frac{\Delta|\Delta|^2}{2\tilde{\omega}_\nu^3}
+\frac{1}{8\tilde{\omega}_\nu^3\tilde{\Omega}_\nu^2}
\left[
	\langle ({\bf vD})^2 \rangle \Delta|\Delta|^2
\right.
\nonumber \\
\left.
	+
	\langle ({\bf vD}) 
			(|\Delta|^2 ({\bf vD}) \Delta 
			+ \Delta^2 ({\bf vD})^* \Delta^* )
	\rangle 
\right.
\nonumber \\
\left.
	+ 
	\Delta 
	\langle 
		|{\bf vD} \Delta)|^2 
	\rangle 
	+
	|\Delta|^2 \langle ({\bf vD})^2 \Delta \rangle
	+\Delta ^2 \langle ({\bf vD}^*)^2 \Delta^* \rangle
\right]
\nonumber \\
+
\frac{1}{16\tau\tilde{\omega}_\nu^4\tilde{\Omega}_\nu^2}
\left[
	\langle ({\bf vD})^2 \rangle \Delta|\Delta|^2
	+
	\langle ({\bf vD})  
			\Delta^2 
			({\bf vD})^* \Delta^* 
	\rangle 
\right.
\nonumber \\
\left.
	+
	|\Delta|^2 \langle ({\bf vD})^2 \rangle \Delta
	+
	\Delta^2 \langle ({\bf vD}^*)^2 \rangle \Delta^*
	\rangle 
\right].
\nonumber \\
\end{eqnarray}

At $\omega_\nu$ of arbitrary sign, one should substitute 
$\tilde{\omega}_\nu \rightarrow |\omega_\nu|-i\mu H 
\text{sign}\omega_\nu$ and $\tilde\Omega_\nu \rightarrow
|\omega_\nu|+1/2\tau-i\mu H \text{sign}\omega_\nu$ in
Eqs.~(\ref{B-f1-Id}), (\ref{B-f3-Id}).

Inserting now Eqs.~(\ref{B-f1-Id}), (\ref{B-f3-Id}) into (\ref{B-gap}), 
we can put the selfconsistency equation for the gap in the form:
\begin{eqnarray} \label{B-gap-Id}
0=
\left(\ln \frac{T}{T_{c0}}
+2 \pi T \sum_{\omega>0} \frac{1}{\omega}-\pi K_{10} \right) \Delta
+
\frac{\pi K_{21}}{4} 
\langle ({\bf vD})^2\rangle \Delta
\nonumber \\
-
\frac{\pi K_{23}}{16}
\langle ({\bf vD})^4\rangle \Delta
-
\frac{\pi K_{33}}{32\tau}
\langle ({\bf vD})^2\rangle^2 \Delta
+
\frac{\pi K_{30}}{2}
\Delta|\Delta|^2
\nonumber \\
-
\frac{\pi K_{32}}{8}
\left[
	4 |\Delta|^2 \langle ({\bf vD})^2 \rangle \Delta
	+
	\Delta^2 \langle ({\bf vD}^*)^2 \rangle \Delta^*
\right.
\nonumber \\
\left.	
	+
	3 \Delta^* \langle ({\bf vD} \Delta) \rangle^2
	-
	2 \Delta |\langle ({\bf vD} \Delta) \rangle|^2 
\right]
\nonumber \\
-
\frac{\pi K_{42}}{16\tau}
\left[
	3 |\Delta|^2 \langle ({\bf vD})^2 \rangle \Delta
	+
	\Delta^2 \langle ({\bf vD}^*)^2 \rangle \Delta^*
\right.
\nonumber \\
\left.
	+
	2 \Delta^* \langle ({\bf vD} \Delta) \rangle^2
	-
	2 \Delta |\langle {\bf vD} \Delta \rangle|^2 
\right],
\nonumber \\
\end{eqnarray}
where we used the standard regularization rule:
\begin{equation}
\frac{1}{V}
=\ln \frac{T}{T_{c0}}
+2 \pi T \sum_{\omega>0} \frac{1}{\omega},
\end{equation}
and the coefficients $K_{nm}$ are defined in Eq.~(\ref{e6}).
We can check straightforwardly that Eq.~(\ref{B-gap-Id}) corresponds 
to the 
saddle point equation for the free energy functional (\ref{e2}):
\begin{equation}
\frac{\delta F}{\delta \Delta^*({\bf r})} = 0.
\end{equation}

\subsection{Non-identity representation}

One should proceed along the same line to derive the gap equation for 
non-identity representation (when $\langle \psi \rangle=0$). In 
particular, one can obtain:
\begin{eqnarray} \label{B-f1-NonId}
\langle 
\psi^* f_\nu^{(1)}
\rangle
=
\frac{\Delta}{\tilde{\Omega}_\nu} 
-\frac{\langle |\psi|^2({\bf vD})^2\rangle 
\Delta}{4\tilde{\Omega}_\nu^3}
+
\frac{\langle |\psi|^2({\bf vD})^4\rangle 
\Delta}{16\tilde{\Omega}_\nu^5}
\nonumber \\
+\frac
	{
		\langle \psi^*({\bf vD})^2\rangle 
		\langle \psi({\bf vD})^2\rangle \Delta
	}
	{32\tau\tilde{\omega}_\nu\tilde{\Omega}_\nu^5}
\qquad
\end{eqnarray}
and
\begin{eqnarray} \label{B-f3-NonId}
\langle 
\psi^* f_\nu^{(3)}
\rangle
=
-\frac{1}{2\tilde{\Omega}_\nu^3} 
\left(
	\langle |\psi|^4  \rangle
	-\frac{1}{2\tau \tilde{\Omega}_\nu}
\right)
\Delta|\Delta|^2
\nonumber \\
+\frac{1}{8\tilde{\Omega}_\nu^5}
\langle |\psi|^4
\left[
({\bf vD})^2 \Delta|\Delta|^2
+
{\bf vD}
	\left(
		|\Delta|^2({\bf vD} \Delta)
\right.
\right.
\nonumber \\
\left.
\left.
		+\Delta^2 ({\bf vD} \Delta)^*
\right)	
 +
	|\Delta|^2({\bf vD})^2 \Delta
\right.
\nonumber \\
\left.	
	+\Delta^2 ({\bf vD}^2 \Delta)^*
	+\Delta |{\bf vD} \Delta|^2
\right]
\rangle
\nonumber \\
-\frac{1}{16 \tau\tilde{\Omega}_\nu^6}
\langle 
	|\psi|^2
	\left[
		({\bf vD}) |\Delta|^2 ({\bf vD}) \Delta
		+
		({\bf vD})^2 |\Delta|^2 \Delta
\right.
\nonumber \\
\left.
		+2|\Delta|^2 ({\bf vD})^2  \Delta
		+\Delta^2 ({\bf vD}^*)^2  \Delta^*
		+\Delta |{\bf vD}  \Delta|^2
	\right]
\rangle.
\nonumber \\
\end{eqnarray}

Inserting now Eqs.~(\ref{B-f1-NonId}, \ref{B-f3-NonId}) into 
(\ref{B-gap}), we can put the selfconsistency equation for the gap in 
the form:
\begin{eqnarray} \label{B-gap-NonId}
0=
\left(
	\ln \frac{T}{T_{c0}}
	+
	2 \pi T 
	\sum_{\omega>0} 
	\frac{1}{\omega}
	-\pi K_{01} 
\right) 
\Delta
\nonumber \\
+
\frac{\pi K_{03}}{4} 
\langle |\psi|^2 ({\bf vD})^2 \rangle \Delta
-
\frac{\pi K_{05}}{16}
\langle |\psi|^4 ({\bf vD})^4\rangle \Delta
\nonumber \\
-
\frac{\pi K_{15}}{32\tau}
\langle \psi^* {\bf vD}\rangle
\langle \psi {\bf vD}\rangle
\Delta
+
\frac{\pi}{2}
\left(
	\langle |\psi|^4 \rangle K_{30}
	-\frac{K_{04}}{2\tau} 
\right)
\Delta |\Delta|^2
\nonumber \\
-
\frac{\pi K_{05}}{8}
\left[
	4 |\Delta|^2 \langle ({\bf vD})^2 \rangle \Delta
	+
	\Delta^2 \langle ({\bf vD}^*)^2 \rangle \Delta^*
\right.
\nonumber \\
\left.	
	+
	3 \Delta^* \langle ({\bf vD} \Delta) \rangle^2
	-
	2 \Delta |\langle ({\bf vD} \Delta) \rangle|^2 
\right]
\nonumber \\
+
\frac{\pi K_{06}}{16\tau}
\left[
	5 |\Delta|^2 \langle ({\bf vD})^2 \rangle \Delta
	+
	2\Delta^2 \langle ({\bf vD}^*)^2 \rangle \Delta^*
\right.
\nonumber \\
\left.
	+
	3 \Delta^* \langle ({\bf vD} \Delta) \rangle^2
	-
	4 \Delta |\langle {\bf vD} \Delta \rangle|^2 
\right].
\nonumber \\
\end{eqnarray}

We can check straightforwardly that Eq.~(\ref{B-gap-NonId}) 
corresponds to the saddle point equation for the free energy 
functionnal (\ref{e7}).

\section{Lowest Landau level approximation}

In this Appendix we shall prove that in the limit of small influence
of orbital effect (large Maki parameters) the function given by Eq.~ 
(\ref{e22}) is appropriate variational function for the Abrikosov
lattice ground state in tetragonal superconductor under magnetic field
directed along c-axis.  With this purpose let us consider the 
Hamiltonian
of the form
\begin{equation}
H=H_{0} +H_{4} ,
\label{eC1}
\end{equation}    
where
\begin{eqnarray}
&&H_{0} =\alpha a^{-} a^{+} + \beta
[a^{-} a^{+} a^{-} a_{+} +a^{-} a^{-} a_{+} a^{+} \nonumber\\
&&+a^{-} a^{+} a^{+} a^{-} +a^{+} a^{-} a^{-} a^{+} \nonumber\\
&&+a^{+} a^{+} a^{-} a^{-} +a^{+} a^{-} a^{+} a^{-} ],
\label{eC2}
\end{eqnarray}
and
\begin{equation}
H_{4} =\gamma (a^{+} )^{4} +\delta (a^{-} )^{4} .
\label{eC3}
\end{equation}
The dimensionless differential operators $a^{\pm} =\lambda D^{\pm} $
act on the Landau states $\phi_{n} (x,y),~~n=0,1,2,\ldots$ as follows
\begin{equation}
a^{-} \phi_{n} =\sqrt {n}\phi_{n-1} ,~~~~a^{+} \phi_{n} =\sqrt 
{n+1}\phi_{n+1} ,
\label{eC4}
\end{equation}
such that
\begin{equation}
H_{0} \phi_{0} =\varepsilon_{0} \phi_{0} ,~~~\varepsilon_{0} =\alpha+3\beta,
\label{eC5}
\end{equation}
and
\begin{equation}
H_{0} \phi_{4} =\varepsilon_{4} \phi_{4} ,~~~\varepsilon_{4} =5\alpha+123\beta.
\label{eC6}
\end{equation}
Let us consider variational wave function
\begin{equation}
\phi=\phi_{0} +a_{4} \phi_{4} 
\label{eC7}
\end{equation}
with $a_{4} $ as a variational parameter and calculate the 
expectation value
\begin{equation}
\frac{\int dxdy \phi H \phi }{\int dxdy|\phi|^{2}  }=
\frac{\varepsilon_{0} +\sqrt{24}(\gamma +
\delta)a_{4} +\varepsilon_{4} a_{4} ^{2} }{1+a_{4} ^{2} }.
\label{eC8}
\end{equation}     
The minimum of this expression is determined as solution of the 
equation
\begin{equation}
(\gamma +\delta)a_{4} ^{2} -
\frac{2(\varepsilon_{4} -\varepsilon_{0} )}{\sqrt{24}}a_{4} -
(\gamma +\delta)=0
\label{eC9}
\end{equation}     
It is clear that at $\gamma =\delta=0$ in other words at $H=H_{0} $
the variational parameter $a_{4} =0$.
In general
\begin{equation}
a_{4} =\nu -\nu\sqrt{1+1/\nu^{2} }\approx -\frac{1}{2\nu},
\label{eC10}
\end{equation} 
where
\begin{equation}
\nu=\frac{\varepsilon_{4} -\varepsilon_{0} }{\sqrt{24}(\gamma 
+\delta)}.
\label{eC11}
\end{equation} 
The values of coefficients we used are
\begin{equation}
\alpha=\frac{\pi 
N_{0} K_{03} \langle|\psi|^{2} v_{\perp} ^{2} \rangle}
{8\lambda^{2} },
\label{eC12}
\end{equation} 
\begin{equation}
\beta =-\frac{\pi 
N_{0} K_{05} \langle|\psi|^{2} v_{\perp} ^{4} \rangle}
{64\lambda^{4} },
\label{eC13}
\end{equation}
and
\begin{equation}
\gamma=\delta=-\frac{\pi N_{0} K_{05} \langle|\psi|^{2} 
(v_{x} ^{4} -6v_{x} ^{2} v_{y} ^{2} +v_{y} ^{4} )\rangle} 
{64\lambda^{4} },
\label{eC14}
\end{equation} 
where, for brevity we have written  $\alpha, \beta, \gamma$ and
$\delta$ in clean limit $\tau\to\infty$).
Thus,in the limit of large $\lambda $
we obtain
\begin{equation}
\nu\approx\frac{16K_{03} \lambda^{2} \langle|\psi|^{2} v_{\perp} ^{2} \rangle|}
{\sqrt {6}(-K_{05} )\langle|\psi|^{2} 
(v_{x} ^{4} -6v_{x} ^{2} v_{y} ^{2} +v_{y} ^{4} )\rangle|}
\label{eC15}
\end{equation} 
Hence, $\nu\propto\lambda^{2} $ and our variational parameter proves 
to
be small as
\begin{equation}
a_{4} =O\left (\frac{\xi_{0} ^{2} }{\lambda^{2} }\right)~\cong ~
O\left (\frac{1}{\alpha_{M} }\right ).
\label{eC16}
\end{equation}


\begin{thebibliography}{99}
    
\bibitem{Gor} L.P.Gor'kov, Zh.  Eksp.Teor.Fiz.  
{\bf 37}, 1407 (1959) [Sov. Phys. JETP {\bf 37}, 998 (1960)].
   
\bibitem{Wert} N.R.Werthamer, E.Helfand and P.C.Hohenberg, 
 Phys.Rev.  {\bf 147}, 295 (1966).

\bibitem{Clog} A.M.Clogston, Phys.Rev.Lett. {\bf 9}, 266
(1962).

\bibitem{Chandr} B.S.Chandrasekhar, Appl.Phys.Lett.  {\bf 1}, 7 
(1962).
 
 
\bibitem{Maki} K.Maki, Physics {\bf 1}, 127 (1964).

\bibitem{Sar} G.Sarma, J. Phys.Chem.Solids {\bf 24}, 1029 (1963).

\bibitem{MaTs} K.Maki and T.Tsuneto, Progr.Theor.Phys.  {\bf 31},
945 (1964).

\bibitem{Saint} D.Saint-James, G.Sarma, E.J.Thomas, {\it Type II
Superconductivity} (Pergamon, New York, 1969).


\bibitem{FF} P.Fulde and R.A.Ferrell, Phys.Rev.  {\bf 135}, A550 
(1964).

\bibitem{LO} A.I.Larkin and Yu.N.Ovchinnikov,  Zh.  Eksp.Teor.Fiz.  
{\bf 47}, 1136 (1964) [Sov. Phys. JETP {\bf 20}, 762 (1965)].

\bibitem{Bow} J.A.Bowers and K.Rajagopal, Phys.Rev.D {\bf 66}, 065002 
(2002).

\bibitem{Buz} A.I.Buzdin and H.Kachkachi, Phys.Lett.  A{\bf 225}, 341
(1997).

\bibitem{Mora} C.Mora and R.Combescot,  Phys.Rev.B {\bf 71}, 214504 
(2005).


\bibitem{Gru} L.W.Gruenberg and L.Gunther,  Phys.Rev.Lett.  {\bf 16}, 
996
(1966)

\bibitem{Asl} L.G.Aslamazov,  Zh.  Eksp.Teor.Fiz.  
{\bf 55}, 1477 (1968) [Sov. Phys. JETP {\bf 28}, 773 (1969)].

\bibitem{Houz}M.Houzet, A.I.Buzdin,  Phys.Rev.B {\bf 63}, 184521 
(2001).

\bibitem{MakiWon} K. Maki and H. Won, Czech J. Phys. {\bf 46}, 1035 
(1996).

\bibitem{Sam}K.V.Samokhin, Physica C {\bf 274}, 156 (1997).

\bibitem{Kun}Kun Yang, S.L.Sondhi, Phys.Rev.B {\bf 57}, 8566 (1998).

\bibitem{Vor}A.V.Vorontsov, J.A.Sauls, M.J.Graf, Phys.Rev.B {\bf 72},
184501 (2005).

\bibitem{Mov} R.Movshovich, M.Jaine, J.D.Thompson, C.Petrovich, 
Z.Fisk,
P.G.Pagliuso, and J.L.Sarrao, Phys.Rev.Lett.  {\bf 86}, 5152 (2001).

\bibitem{Iza} K.Izawa, H.Yamaguchi, Yu.Matsuda, H.Shishido, R.Settai,
and Y.Onuki, Phys.Rev.Lett.  {\bf 87}, 057002 (2001).

\bibitem{Bian02}A.Bianchi, R.Movshovich, N.Oeschler, P.Gegenwart, 
F.Steglich,
J.D.Thompson, P.G.Pagliuso, and J.L.Sarrao,  Phys.Rev.Lett.  {\bf 
89}, 137002
(2002).

\bibitem{Mic} C. F. Miclea, M. Niclas, D. Parker, K. Maki, J. L. 
Sarrao, 
J. D. Thompson, G. Sparn, and F. Steglich, Phys. Rev. Lett. {\bf 96}, 
117001
(2006).

\bibitem{Bian03}A.Bianchi, R.Movshovich,
C.Capan, P.G.Pagliuso, and J.L.Sarrao, Phys.Rev.Lett.  {\bf 91}, 
187004
(2003).

\bibitem{Agt} D. F. Agterberg and K. Yang, J. Phys. Condens. Matter 
{\bf 
13}, 9259 (2001).

\bibitem{Ik} H. Adachi, R. Ikeda, Phys.Rev.B {\bf 68}, 184510 (2003);
R. Ikeda, H. Adachi, Phys.Rev.B {\bf 69}, 212506 (2004).

\bibitem{Ovc} Yu. N. Ovchinnikov,  Zh.  Eksp.Teor.Fiz.  
{\bf 115}, 726 (1999) [JETP {\bf 88}, 398 (1999)].

\bibitem{Eil} G. Eilenberger, Z.Phys.{\bf 214}, 195 (1968).

\bibitem{LaOv} A. I. Larkin and Yu. N. Ovchinnikov, Zh.  
Eksp.Teor.Fiz.  
{\bf 55}, 2262 (1968) [Sov. Phys. JETP {\bf 28}, 1200 (1969)].

\bibitem{Gor60} L. P. Gor'kov, Zh.  Eksp.Teor.Fiz.  {\bf 37}, 1407  
(1959) [Sov.  Phys.  JETP {\bf 10}, 998 (1960)].

\bibitem{Min} V. P. Mineev and K. V. Samokhin, "Introduction to
nonconventional superconductivity", Gordon and Breach Sc.  Publ., OPA 
1999.

\bibitem{Abrikosov} A. A. Abrikosov, {\it Fundamentals of the theory 
of metals},
(North-Holland New York, NY, 1988). 

\bibitem{BuzBris} A. I. Buzdin and J. P. Brison, Phys. Lett. A {\bf 
218}, 359 (1996).


\bibitem{Vav} M. G. Vavilov and V. P. Mineev, Zh.  Eksp.Teor.Fiz.  
{\bf 113}, 
2174 (1998) [JETP {\bf 86}, 1191 (1998)].

\bibitem{maki66} K. Maki, Phys. Rev. {\bf 148}, 362-369 (1966).

\bibitem{K.Maki} K.Maki, Physics {\bf 1}, 127 (1964).

\bibitem{kopnin} N. Kopnin, {\it Theory of nonequilibrium 
superconductivity} (Clarendon Press, Oxford, 2001).






\end{thebibliography}
\end{document}